\documentclass{aa}  

\pdfoutput=1
\usepackage{graphicx}
\usepackage{subcaption}
\usepackage{txfonts}
\usepackage{amssymb}
\usepackage{xcolor,colortbl}
\usepackage{graphicx}

\usepackage{array}
\usepackage{lscape}
\usepackage{rotating}
\usepackage{longtable}
\usepackage{color}
\usepackage{amsfonts}

\newcommand{\teff}{$T_{\mathrm{eff}}$~}
\newcommand{\logg}{$\mathrm{log}\,g$~} 
\newcommand{\asini}{$a\,\mathrm{sin}\,i$~} 
\newcommand{\vsini}{$v\,\mathrm{sin}\,i$~} 
\newcommand{\teffi}{$T_{\mathrm{eff}}$}
\newcommand{\loggi}{$\mathrm{log}\,g$} 
\newcommand{\vsinii}{$v\,\mathrm{sin}\,i$} 

\newcommand{\kms}{km\,s$^{-1}$~}
\newcommand{\kmsi}{km\,s$^{-1}$}

\newcommand{\porb}{$P_{\mathrm{orb}}$~}
\begin{document}

   \title{Orbital solutions derived from radial velocities and time delays for four {\it Kepler} systems with A/F-type {(candidate)} hybrid pulsators\thanks{This 
   study is based on spectra obtained with the {HERMES} \'echelle spectrograph installed at the Mercator telescope, operated by the IvS, KULeuven, 
   funded by the Flemish Community and located at the Observatorio del Roque de los Muchachos, La Palma, Spain of the Instituto de Astrof\'{\i}sica 
   de Canarias. 
   } } 
   \author{P. Lampens\inst{1}
          \and
           L. Vermeylen\inst{1}
          \and
           Y. Fr\'emat\inst{1} 
          \and
           \'A. S\'odor\inst{2}
          \and
           M. Skarka\inst{3,4}
          \and
           A. Samadi-Ghadim\inst{5} 
          \and
           Zs. Bogn\'ar\inst{2,6}
          \and
           H. Lehmann\inst{7}  
          \and
           P. De Cat\inst{1}
          \and
           A. Goswami\inst{8}
          \and
           L. Dumortier\inst{1}
           }
   \institute{Koninklijke Sterrenwacht van Belgi\"e, Ringlaan 3, 1180 Brussel, Belgium\\
              \email{patricia.lampens@oma.be}
          \and
   Konkoly Observatory, Research Centre for Astronomy and Earth Sciences \& MTA CSFK Lend\"ulet Near-Field Cosmology Research Group, Konkoly Thege M. u. 15-17, H-1121, Budapest, Hungary
          \and
   Astronomical Institute, Czech Academy of Sciences, Fri\v cova 298, CZ-25165 Ond\v rejov, Czech Republic
          \and
   Department of Theoretical Physics and Astrophysics, Masaryk Univerzity, Kotl\'a\v rsk\'a 2, CZ-61137 Brno, Czech Republic
          \and
   N\'ucleo de Astronom\'{\i}a, Faculdad de Ingenier\'{\i}a y Ciencias, Universidad Diego Portales, Av. Ej\'ercito Libertador 441, Santiago, Chile
          \and
   ELTE E\"otv\"os Lor\'and University, Institute of Physics, P\'azm\'any P\'eter s\'et\'any 1/A, H-1171, Budapest, Hungary
          \and
   Th\"uringer Landessternwarte, Tautenburg, Germany
          \and
   Indian Institute of Astrophysics, II Block, Koramangala, Bangalore 560 034, India
         }

   \date{Received...; accepted ...}

 
  \abstract
   {The presence of  A/F-type {\it Kepler} hybrid stars extending across the entire $\delta$ Sct-$\gamma$ Dor instability strips
and beyond remains largely unexplained. In order to better understand these particular stars, we performed a multi-epoch spectroscopic 
study of a sample of 49 candidate A/F-type hybrid stars and one cool(er) hybrid object detected by the {\it Kepler} mission.
We determined a lower limit of 27\% for the multiplicity fraction. For six spectroscopic systems, we also reported 
long-term variations of the time delays. For four systems, the time delay variations are fully coherent with those of 
the radial velocities (RVs) and can be attributed to orbital motion. }
   {We aim to improve the orbital solutions for those spectroscopic systems with long orbital periods (order of 4-6 years) 
among the {\it Kepler} hybrid stars that we continued to observe. }
   {The orbits are computed based on a simultaneous modelling of the RVs obtained with high-resolution spectrographs 
and the photometric time delays (TDs) derived from time-dependent frequency analyses of the {\it Kepler} light curves. 
   }
   {We refined the orbital solutions of four spectroscopic systems with A/F-type {\it Kepler} hybrid component stars: KIC~4480321,
~5219533, ~8975515 and KIC~9775454. Simultaneous modelling of both data types analysed together enabled us to improve 
the orbital solutions (all), obtain more robust and accurate information on the mass ratio (some for the first time), 
and identify the component with the short-period $\delta$ Sct-type pulsations (all). 
The information gained is maximized when one of the components, generally the one exhibiting the $\delta$ Sct-type pulsations, 
is the fast(er) spinning component. In several cases, we were also able to derive new constraints for the minimum component masses. 
{From a search for regular frequency patterns in the high-frequency regime of the Fourier transforms of each system, 
we found no evidence of tidal splitting among the triple systems with close (inner) companions. However, some systems exhibit 
frequency spacings which can be explained by the mechanism of rotational splitting.} }
   {}


\titlerunning{Orbital solutions for four {\it Kepler} systems with hybrid pulsations}
\authorrunning{Lampens et al.}

   \maketitle
%

\section{Introduction}

In an effort to understand and unravel the enigma of the low frequencies in the brighter A/F-type candidate hybrid pulsators of the 
{\it Kepler} mission \citep{Grigahcene2010ApJ...713L.192G,Uytterhoeven2011A&A...534A.125U,Balona2015MNRAS.452.3073B,Bowman2017ampm.book.....B}, 
we started a radial-velocity monitoring campaign with high-resolution \'echelle spectrographs located in various European observatories. 
Our goal is to characterize the spectroscopic variability of an unbiased sample of {\it Kepler} hybrid ($\gamma$ Dor - $\delta$ Sct) pulsators. 
This programme was initiated during the middle of 2013. We collected multi-epoch observations (4 - 6 times at least) in order to detect 
binarity or multiplicity at the different time scales, with orbital periods ranging from a few days to several years, and to establish 
a meaningful classification. We determined the epoch radial velocities, projected rotational velocities, new or improved atmospheric 
stellar properties (i.e. \teffi, \loggi, \vsinii), and we provided a classification in terms of multiplicity, pulsation and/or 
(fast) rotation for all our targets on the basis of the shapes of the cross-correlation functions and their radial velocities as a 
function of time \citep[][from hereon Paper~I]{Lampens2018A&A...610A..17L}. \\ 

In this first study, we followed a sample of 49 candidate hybrid stars and one much cooler hybrid object identified by 
\citet{Uytterhoeven2011A&A...534A.125U}. 
We found a significant rate of short- and long-period binary as well as multiple systems as we detected 10 spectroscopic systems in total, 
i.e. 4 double-lined (SB2) systems, 3 triple-lined (SB3) systems, 4 single-lined (SB1) systems (only 3 belong to the A/F-class), 
and 3 objects with long-term radial-velocity variations (VAR). We determined the orbital solutions of seven systems. For two hierarchical 
triple systems, we also proposed a preliminary solution for the outer orbit. Using the classification results, we provided a lower limit 
of the fraction of A/F-type hybrid pulsators which belong to spectroscopic binary and multiple systems. Including the known {\it Kepler} 
eclipsing binary KIC~11180361 (KOI-971), we derived a minimum multiplicity fraction of 27\%. Two other hybrid targets have a possible 
companion or shell (CMP). If we count these two targets, we obtain an overall multiplicity fraction of about 30\% among these candidate 
hybrid pulsators. \\

Among the new spectroscopic systems, we identified {four} cases for which analysis of the {\it Kepler} time delays (TDs) (mostly 
derived from the short-period pulsation frequencies) with the radial velocities (RVs) enables us to derive improved orbital elements, 
accurate mass ratios as well as a (most) probable identification of the pulsating component. The goal of this paper is to determine 
orbital solutions for these systems as accurately as possible based on the results of a unified modelling by combining the various 
data types into a single, simultaneous analysis. In the present study, we will make usage of the radial velocities acquired with the 
\'echelle spectrograph HERMES equipping the 1.2-m Mercator telescope, La Palma, Spain \citep{Raskin2011A&A...526A..69R} 
available from Paper~I (see table C.2). Furthermore, we reported variable TDs in nine cases, 
in particular for the spectroscopic systems KIC~4480321, 5219533, 8975515, and KIC~9775454 (Paper~I, fig.~24). 
For all four systems, our spectroscopic observations were acquired until late 2019. For KIC~9775454, we included the measurements obtained 
in 2018 with the high-resolution \'echelle spectrograph {HESP} (R = 60,000) mounted at the Himalayan Chandrasekhar telescope (HCT), 
operated by the IIA, Bangalore, India\footnotemark. Consequently, all RV data of all four systems and their components were updated 
to the latest possible date. First, we present the current status of each individual system (Sect.~\ref{Indiv}). In 
Sect.~\ref{themethod}, we introduce the methodology for computing the refined orbits. Sect.~\ref{orb:combined} provides the 
orbital solutions using simultaneous modelling. 
In Sect.~\ref{sec:frequent_spacings}, we present and discuss the distributions of the frequency spacings in the high-frequency region 
of the Fourier transforms of each system. Finally, we present our conclusions for all the treated systems (Sect.~\ref{sec:conclusions}). 
\footnotetext{\hbox{https://www.iiap.res.in/hesp}}\\

\section{Presentation of the systems}\label{Indiv}

We present a summary of useful high-resolution spectra for each of the four systems listed in Table~\ref{tab:log}. In addition to the total 
number of spectra available for each system, we also mention the spectrograph and the time basis. The revised orbital solutions 
(Sect.~\ref{orb:combined}) are thus based on the combination of updated sets of component RVs and the time delays (TDs) previously derived 
from time-dependent frequency analyses of the oscillations detected in the {\it Kepler} light curves (Paper~I), providing us with a total 
time basis longer than eight years for each case. \\   

\setlength\tabcolsep{3pt}
\renewcommand{\arraystretch}{1}
\begin{table}
\centering
\caption{\label{tab:log} List of targets and general properties of the RV data.}		
\begin{tabular}{ccccc@{}c}
\hline
\hline 
KIC ID 	& Nr & BJD start & BJD end & Time range & Instr.\footnotemark\\ 
 	&    &{\small 2,450,000.+} & {\small 2,450,000.+}  &  (from/to) & \\ \hline
{\small 4480321} & {\small 61}	& {\small 5820.4499} & {\small 8676.4431} &  {\tiny 15/09/2011 - 11/07/2019}& {\small H1} \\
{\small 5219533} & {\small 52}	& {\small 5372.7083} & {\small 8792.3578} &  {\tiny 25/06/2010 - 04/11/2019}& {\small H1} \\
{\small 8975515} & {\small 31}	& {\small 5345.7088} & {\small 8792.3686} &  {\tiny 29/05/2010 - 04/11/2019}& {\small H1} \\
{\small 9775454} & {\small 26}	& {\small 5345.7325} & {\small 8794.3056} &  {\tiny 29/05/2010 - 06/11/2019}& {\small H1} \\ 	
{\small 9775454} & {\small  2}	& {\small 8231.4431} & {\small 8232.3771} &  {\tiny 22/04/2018 - 23/04/2018}& {\small H2} \\ 	
\hline
\end{tabular}
\tablefoot{\tablefoottext{2}{H1 stands for 'HERMES'. H2 stands for 'HESP'.}}\\
\end{table}

\subsection{KIC~4480321}\label{KIC4480321}

KIC~4480321 (HD~225479, V=10.3, A9~Vwkmet) is a hybrid variable star revealed by \citet{Uytterhoeven2011A&A...534A.125U}, 
with frequencies in the ranges [0.2, 5.0] d$^{-1}$ and [5.1, 61.2] d$^{-1}$ and a most dominant frequency of 0.710 d$^{-1}$ 
in the $\gamma$ Dor region. The star was also reported as rotationally variable \citep{Nielsen2013A&A...557L..10N}, 
with g-mode multiplets split by rotation. The period spacings in the $\gamma$ Dor region were obtained by \citet{2019MNRAS.482.1757L}. 
This star was classified as {an} SB3 from our multi-epoch study. 
It is {a} hierarchical system, with a twin-like inner binary consisting of early F-type stars orbiting a slightly more luminous and more rapidly rotating A-type 
companion (we used the basic model of an A5-star, \vsini = 160 \kms with 2 F0-stars, \vsini = 10 \kmsi, Paper~I). We derived an orbital solution 
for the inner pair with a 9.1659 d period, and an outer orbital solution with an estimated 2280 d period, and reported the existence of variable 
TDs in excellent agreement with the proposed solution for the outer orbit. We determined the component effective temperatures \teffi$_{1}$ = 7900 $\pm$ 100 K 
and \teffi$_{2}$ and \teffi$_ {3}$ ranging between 6300 and 6900 K (the latter components have almost similar temperatures), assuming \logg = 4~dex (cgs) 
(Paper~I, Sect.~6.3). \citet[][{Appendix C3}]{Murphy2018MNRAS.474.4322M} 
reported $P$ = 2270 $\pm$ 60 d for the outer orbit from a modelling of \citet{Lampens2018A&A...610A..17L}'s RV measurements in combination with their 
time delays. \\

\subsection{KIC~5219533}\label{KIC5219533}

KIC~5219533 \citep[HD~226766, BU~1474~B, V=9.2, A2-A8,][]{Renson2009A&A...498..961R} is the visual companion of BU~1474~A at an angular separation 
of 65" (HD~189178, V=5.44, B5 He weak \citep{Renson2009A&A...498..961R}; also a spectroscopic binary). This star was classified as a {\it Kepler} 
hybrid variable star by \citet{Uytterhoeven2011A&A...534A.125U}, with frequencies in the ranges [0.3, 4.6] d$^{-1}$ and [5.4, 29.9] d$^{-1}$, 
and a dominant frequency of 10.285 d$^{-1}$ \citep[][Table 3]{Uytterhoeven2011A&A...534A.125U}. 
It was recognized as a new SB3 based on the multi-epoch spectra. 
The system is hierarchical, consisting of an inner pair with two nearly identical stars of spectral type near A5 and a more rapidly rotating 
outer component of a slightly cooler type. For both components of the inner pair, the mean effective temperatures of \teffi$_{1}$ = 8300 
$\pm$ 100 K and \teffi$_{2}$ = 8200 $\pm$ 100 K and projected rotational velocities of 10 \kms were obtained coupled to the light factor 
l$_{1}$ = 0.53 $\pm$ 0.02, assuming \logg = 4~dex (cgs)(Paper~I, Sect.~6.4). Component C has \vsini equal to 115 \kms \citet{Uytterhoeven2011A&A...534A.125U}. 
An orbital solution with \porb = 31.9181 d was derived for the inner pair, while a tentative outer orbit was proposed with 
\porb $\sim$ 1600 d. Later on, \citet[][{Appendix C4}]{Murphy2018MNRAS.474.4322M} 
studied this system using the phase modulation (PM) method \citep{Murphy2014MNRAS.441.2515M}. 
They derived a very high mass function and proposed that the third body is a $\delta$ Sct star with an orbital period longer than 1500 d. 
Simultaneously with \citet{Lampens2018A&A...610A..17L}, \citet{Catanzaro2019MNRAS.487..919C,Catanzaro2019MNRAS.488..480C} 
independently observed the system (from 2014 to 2018) with the aim to characterize the orbits, and to perform a chemical analysis of its 
components. They proposed the orbital parameters based on their and \citet{Lampens2018A&A...610A..17L}'s RVs ($P_{\rm 1}$ = 31.9187 d, 
$e_{\rm 1}$ = 0.28, $q_{\rm 1}$ = 1.03 and $P_{\rm 2}$ = 1615 d, $e_{\rm 2}$ = 0.54, $f(M_{\rm 1,2}$) = 0.18) which confirm the findings of Paper~I. 
They also derived the atmospheric properties \teffi, \loggi, and \vsini based on 2 spectra (obtained at max RV separation) and chemical 
abundances based on one spectrum (max S/N), first for two components, next with the third one included. However, they were unable 
to derive the chemical composition of component C (except for Mg II). They concluded that both components of the close pair are twin 
Am stars (with underabundances of C, O, Mg, Ca and Sc and overabundances of Na, Fe-peak elements). From the simple assumption that all 
the components have equal mass, they suggested that the orbits might be coplanar. All previous studies explicitly mention that more data 
for this multiple system are needed. \\    

\subsection{KIC~8975515}\label{KIC8975515}

KIC~8975515 (HD~188538, V=9.5, A6~V:) was classified as a {\it Kepler} hybrid star by \citet{Uytterhoeven2011A&A...534A.125U}. The star shows 
frequencies in the ranges [0.3 - 4.7] d$^{-1}$ and [5.3 - 25.8] d$^{-1}$ of the $\gamma$ Dor and the $\delta$ Sct pulsation regimes respectively, 
with the dominant frequency of 13.97 d$^{-1}$ \citep[][Table 3]{Uytterhoeven2011A&A...534A.125U}. 
Therefore, we included it among our sample of A/F-type candidate hybrid stars to be monitored spectroscopically where it was recognized as a new 
SB2 from the multi-epoch observations. 
The system consists of two A-type stars of similar temperature but with dissimilar projected rotational velocities, forming an eccentric system 
($e$ = 0.409 $\pm$ 0.015). From a detailed spectrum fitting using the code SYNSPEC \citep{Hubeny1995ApJ...439..875H} and the ATLAS-9 atmosphere 
models \citep{Castelli2003IAUS..210P.A20C} in order to build a suitable composite model in the spectral range [500-520] nm, the mean 
effective temperatures \teffi$_{1}$ = 7440 $\pm$ 20 K and \teffi$_{2}$ = 7380 $\pm$ 21 K with the light factor $l_{\rm 1}$ 
of 0.65 $\pm$ 0.03 (component~A), assuming \logg = 4~dex (cgs), were obtained. The mean projected rotation velocities were consistently determined as 
162 $\pm$ 2 and 32 $\pm$ 1 \kms (Paper~I, Sect.~6.12). \citet{Lampens2018A&A...610A..17L} proposed a preliminary orbit with a period of the order 
of 1600 d, in line with the TD curve, while \citet[][{Appendix C12}]{Murphy2018MNRAS.474.4322M} 
obtained a solution with a shorter period of 1090 d, 
using the published RVs in combination with their time delays (TDs). They furthermore assumed that the narrow-lined component is the $\delta$ 
Scuti pulsator and reported a slight aperiodicity in their derived TDs. \\
An extensive study of its pulsational properties was recently published by \citet{Samadi2020A&A...638A..57S}. They concluded that 
both components are pulsating, the fast-rotating component is a pulsating hybrid star also showing retrograde r modes, while the 
more slowly-rotating component is a $\delta$ Scuti pulsator.\\

\subsection{KIC~9775454}\label{KIC9775454}

KIC~9775454 (HD~185115, V=8.2, F1~Vs) is a candidate hybrid variable star from \citet{Uytterhoeven2011A&A...534A.125U}, with frequencies in the 
ranges [0.2, 4.6] d$^{-1}$ and [14.7, 14.9] d$^{-1}$ and a dominant frequency of 4.161 d$^{-1}$. It has a Gaia DR2 RV of -20.22 $\pm$ 1.02 {\kms}
\citep{2018yCat.1345....0G}. 
The atmospheric parameters derived from high-resolution spectroscopy are \teff = 7287 K and \logg = 4.25~dex (cgs) with a projected rotation velocity of
\vsini = 65 \kms (Paper~I, Table C.1). \citet{Lampens2018A&A...610A..17L} classified this object as a long-term SB1 of unknown orbital period. 
The {cross-correlation function} (CCF) of the composite spectrum shows a broad-lined component blended with a narrow-lined component. \citet[][{Appendix C14}]{Murphy2018MNRAS.474.4322M} 
proposed an orbit with \porb = 1686 $\pm$ 13 d and $e$ = 0.23 $\pm$ 0.02 by combining the published RVs with their TDs.  
In this case, the secondary component contributes only weakly to the composite spectrum. Nevertheless, we were able to obtain some individual RVs 
for this (much) cooler component (cf. Sect.~\ref{cas:KIC9775454}). \\

\section{Methodology}\label{themethod}

The orbital motion of a binary system causes a periodic fluctuation of the light path and its associated travel time with respect to the system's centre of mass. 
On the other hand, light from a source in motion undergoes a (Doppler) shift of its frequency. In the frequency domain (for a time base $T$ >> \porb), this 
phenomenon corresponds to a frequency modulation (FM) which can be detected as a frequency multiplet whose exact spacing is the orbital frequency \citep{Shibahashi2012MNRAS.422..738S}. 
In the time domain (for $T$ << \porb), this effect corresponds to a modulation of the phase (PM) with the orbital period, whose amplitude depends on the mass 
of the (unseen) companion and the observed frequency \citep{Murphy2014MNRAS.441.2515M}. Hence, if one of the components is a pulsating star, the change in the 
light travel time introduces periodic phase shifts of each individual (pulsation) frequency. By converting the phase shift into a time delay, 
the frequency dependency is removed. Therefore, if the star is a multiperiodic pulsator, all the pulsation frequencies will experience the same time delay. \\
\citet{Murphy2014MNRAS.441.2515M} described the principle of obtaining time delays from observed phase shifts of the pulsations and applied the PM method 
unto various cases of {\it Kepler} pulsating stars in (non-eclipsing) binary systems, while \citet{Murphy2015MNRAS.450.4475M} 
provided an analytical method for solving the orbit, also in highly eccentric cases. This method is in essence similar to the search for variability in the 
(O-C)'s, the residuals in the times of specific phases in the light curve of periodically pulsating stars which are members of a binary system
\citep[e.g.][]{Moffett1988AJ.....95.1534M, Fauvaud2010A&A...515A..39F}. 
\citet{Murphy2016MNRAS.461.4215M} presented examples of orbital analyses combining radial velocities (RVs) with photometric time delays (TDs), 
and discussed some of the limitations and advantages. 
In this study, we developed and applied our own code for a simultaneous least-squares fitting of our updated RVs and the TDs based on the {\it Kepler} 
data. Both quantities are functions of the same orbital parameters: the time delay corresponds to the light travel time and the radial velocity is the 
time derivative of the position along the orbital path projected in the observer's direction. We consider the following equations as our model:\\
\begin{equation}
\mathrm{TD}(t) = \frac{a\,\mathrm{sin}\,i}{c}\frac{1-e^2}{1+e\,\mathrm{cos}\,(\nu(t))} \mathrm{sin}\,(\nu(t)+\omega)
\label{eq:TD}
\end{equation}

\begin{equation}
V(t) = 2\pi\frac{a\,\mathrm{sin}\,i}{P\sqrt{1-e^2}}(\mathrm{cos}\,(\nu(t)+\omega)+e\,\mathrm{cos}\,\omega) + V_{\gamma},
\label{eq:RV}
\end{equation}

\noindent where the {true anomaly} $\nu$ is a function of $t$, $T_0$, $P$ and $e$, $T_0$ is the time of periastron passage, $P$ is the orbital period, 
\asini is the projected semi-major axis, $e$ is the eccentricity, $\omega$ is the longitude of the periastron, 
$V_{\gamma}$ is the systemic velocity {and c is the speed of light}. Equ.~\ref{eq:TD} is given by \citet{Irwin1952ApJ...116..211I} \citep[see][]{Pribulla2005ASPC..335..103P}. 
{We removed a linear trend from the original TDs computed with Equ.~\ref{eq:TD} \citep{Murphy2016MNRAS.461.4215M} in order to convert the pulsation 
frequencies to their intrinsic values, i.e. unaffected by the variable light travel time of the pulsating component. This correction is necessary} since the 
frequencies used for computing the TDs may be biased by the (uneven) orbital phase coverage of the {\it Kepler} data, thereby introducing a {bias} 
in the TDs themselves (as in the (O-C) curves). \\ 

In Paper~I, we computed the time-dependent phases for up to 20 frequencies of highest $S/N$ (generally the ones located in the $p$-mode regime) for all 
49 targets of our sample. We thus uncovered nine objects for which correlated variations of comparable amplitude in the TDs were observed. In six systems, 
we could attribute the TD variations to orbital motion of a binary or multiple system. By analysing the RV and the TD data together, we are able to refine 
the orbital solutions and obtain stronger constraints on (some of) the orbital parameters and their derived parameters. Since the TDs are generally
determined from the high(er) frequencies, they only provide information on the origin of the $\delta$ Scuti-type pulsations associated to these 
frequencies. However, {it is much harder to identify the origin of the low(er) frequencies ($\gamma$ Doradus-type) due to their longer periods}. 
In the next sections, we will analyse each of the four systems just described and highlight the gains that may be obtained from such modelling. \\

\section{Orbital solutions from combined modelling}\label{orb:combined}

\subsection{The triple system KIC~4480321}\label{cas:KIC4480321}

\begin{figure*}[ht]
 	\centering
  \includegraphics[width=0.80\linewidth]{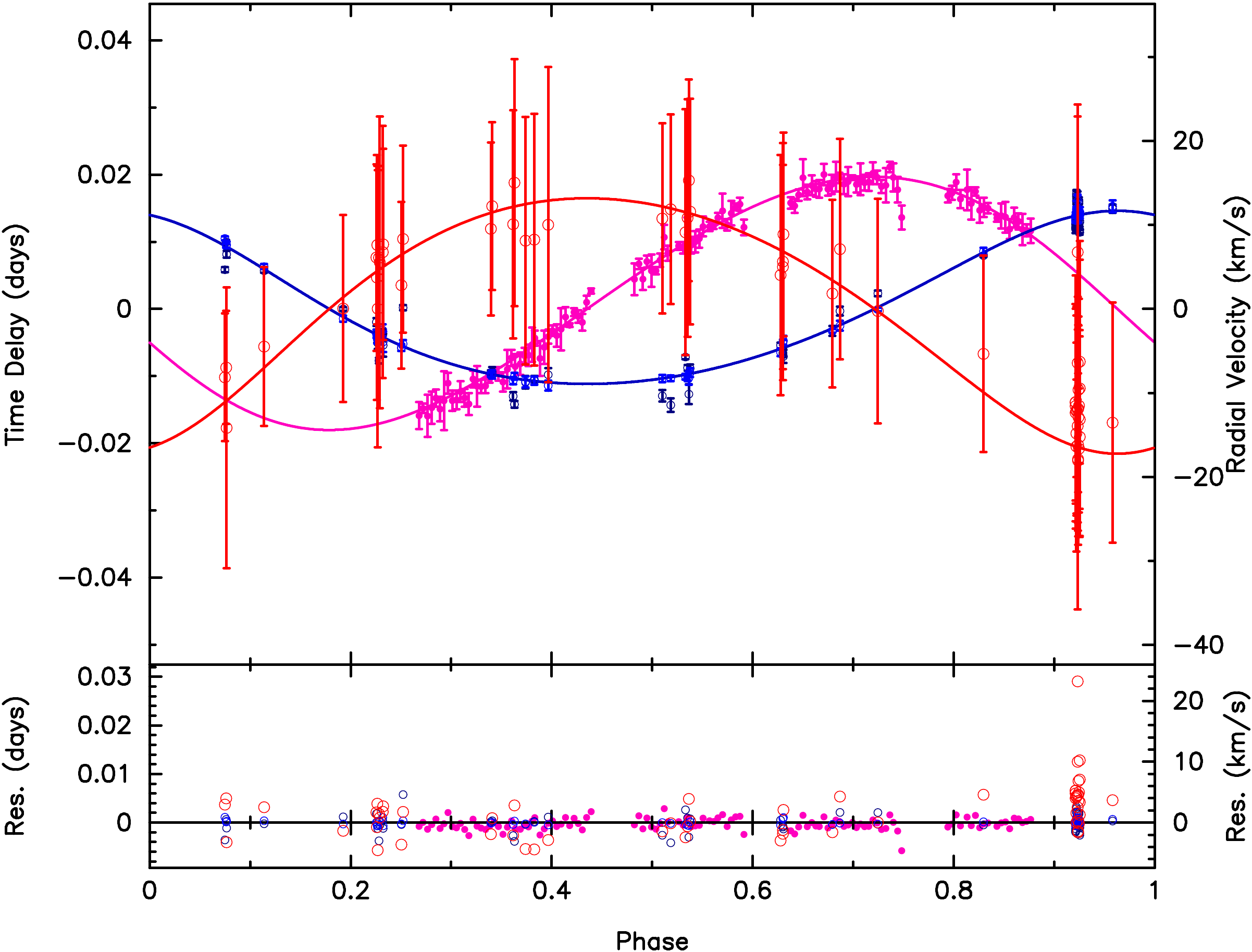}
	\caption{Data and outer orbital solution for KIC~4480321. The {TDs and RVs} (resp. filled pink and unfilled red symbols for component C; 
  blue symbols for the centre of mass of components A and B) are overlaid with the AB-C orbital solution (solid lines). 
  The residuals are shown in the bottom panel. }
	\label{fig:fig_KIC448}
\end{figure*}

KIC~4480321 is a triple system (SB3) whose orbital periods are estimated to be 9.166 and $\sim$ 2280 d (Sect.~\ref{KIC4480321}). We computed the phase 
shifts and their associated TDs for frequencies of highest $S/N$ and evidenced the presence of long-term variations which are coherent with the long-term 
RV variability (Paper~I, fig.~24). We remark that the TDs were corrected by removing a linear trend from the original data (as everywhere). We then 
performed a simultaneous least-squares fitting of three updated component RVs and the mean TDs with their respective uncertainties. The best 
fit in the sense of {min} $\chi^2$ was obtained when we associated the TDs with the RVs of the fast spinning, more luminous A-type star (component C). 
Fig.~\ref{fig:fig_KIC448} illustrates the quality of the fit showing the outer orbital solution together with both data types. The TDs and the RVs 
are overlaid with the AB-C orbital solution. Note that the TD curve mimics the RV curve but with an apparently larger eccentricity and a 90\degr 
shift in periastron longitude \citep{Irwin1952ApJ...116..211I}. 
The revised orbital parameters of the coupled solutions are listed in Table~\ref{tab:KIC448}. Compared to Paper~I, we see that the orbital solution of the 
A-B system is robust and well-determined while the outer orbital solution changed profoundly. The orbital period is longer (\porb $\sim$ 2380 d) and better 
determined than before \citep[see also][]{Murphy2018MNRAS.474.4322M}, 
while the eccentricity and the semi-axis major increased substantially. Thus, the minimum component mass $M_{\rm{C}}\,\mathrm{sin}^{3}\,i_{\mathrm{out}}$ increased 
to 1.60 $\pm$ 0.05 M$_{\odot}$. 
The strongest improvement concerns the accuracy and robustness of the outer orbital solution. This is due to two things: (a) by combining 
both data types, we dispose of a time base longer than eight years and (b) the TDs concern the fast-rotating component C for which the RVs show 
the largest scatter. 
We therefore were able to determine a (much) improved mass ratio for the outer system and to provide (more) reliable values of the three minimum 
component masses. From the minimum total mass of the A-B system derived for each orbit, we get
sin\,$i$ = {0.846 ($\pm$~0.028)} sin\,$i_{\mathrm{out}}$, from which we obtain the condition $i <$ 58\degr~for the inner orbit. 
We may furthermore attribute the $\delta$ Scuti pulsations to the more massive component~C (the frequencies showing phase shifts are all 
located in the $p$-mode region). In summary, KIC~4480321 is a hierarchical triple system with $q_{\mathrm{in}}$ = 
{$M_{\rm{B}}/M_{\rm{A}}$ = 0.98} and $q_{\mathrm{out}}$ = {$M_{\rm{C}}/M_{\rm{AB}}$ = 0.68}.\\

\begin{table}
\center
\begin{minipage}{8.9cm}
\centering \caption[]{\label{tab:KIC448} Values and standard deviations of the constrained orbital parameters for KIC~4480321. Note that from hereon, for 
all similar tables, $T_{\rm{0}}$ is expressed in Hel. JD - 2,400,000. {and $rms$ indicates the standard deviation of the squared residuals.} }
\begin{tabular}{lr@{}l@{}lr@{}l@{}l}
\hline
\hline
\multicolumn{7}{c}{Orbital solution A-B}\\
\hline
Orbital parameter &\multicolumn{3}{c}{Value} & \multicolumn{3}{c}{Std. dev.}\\
\hline
$P$ (d)&9&.&16585 &0&.&00015\\
$T_{\rm{0}}$ (Hel. JD)& 58503&.&065 &0&.&015\\
$e$  &0&.&07894 &0&.&00097\\
$\omega$ (\degr)&351&.&1 &0&.&6\\
$V_{\gamma}$ (\kmsi)& (var.)& & & & &\\
$a_{\rm{A}} \mathrm{sin}\,i$ (au) &0&.&04779 &0&.&00006\\
$a_{\rm{B}} \mathrm{sin}\,i$ (au) &0&.&04876 &0&.&00006\\
\hline
$K_{\rm{A}}$ (\kmsi) & 56&.&90 & 0&.&07\\
$K_{\rm{B}}$ (\kmsi) & 58&.&05 & 0&.&07\\
$M_{\rm{A}} \mathrm{sin}^{3}\,i$ (M$_\odot$) &0&.&7216 &0&.&0019\\
$M_{\rm{B}} \mathrm{sin}^{3}\,i$ (M$_\odot$) &0&.&7073 &0&.&0019\\
$q_{\rm{in}}$ &  0&.&980 & 0&.&002  \\
\hline
$rms_{\rm{A}}$ (\kmsi) & 1&.&486 & & &\\
$rms_{\rm{B}}$ (\kmsi) & 0&.&455 & & &\\
\hline
\hline
\multicolumn{7}{c}{Orbital solution AB-C}\\
\hline
Orbital parameter &\multicolumn{3}{c}{Value} & \multicolumn{3}{c}{Std. dev.}\\
\hline
$P$ (d)&2381&.& &5&.&9\\
$T_{\rm{0}}$ (Hel. JD)& 59082&.& &13&.&\\
$e$  &0&.&140 &0&.&006\\
$\omega$ (\degr)&197&.&9 &1&.&8\\
$V_{\gamma}$ (\kmsi)&-19&.&97 & 0&.&04\\
$a_{\rm{AB}} \mathrm{sin}\,i_{\mathrm{out}}$ (au) &2&.&232 &0&.&013\\
$a_{\rm{C}} \mathrm{sin}\,i_{\mathrm{out}}$ (au) &3&.&293 &0&.&076\\
\hline
$a_{\rm{TD}_{\rm{C}}}/c\, \mathrm{sin}\,i_{\mathrm{out}}$ (d) & 0&.&0190 & 0&.&0004\\
$K_{\rm{AB}}$ (\kmsi) & 10&.&30 & 0&.&07\\
$K_{\rm{C}}$ (\kmsi) & 15&.&20 & 0&.&35\\
$M_{\rm{AB}} \mathrm{sin}^{3}\,i_{\mathrm{out}}$ (M$_\odot$) &2&.&36 &0&.&12\\
$M_{\rm{C}} \mathrm{sin}^{3}\,i_{\mathrm{out}}$ (M$_\odot$) &1&.&60 &0&.&05\\
$q_{\rm{out}}$ &  0&.&678 & 0&.&016 \\
\hline
$rms_{\rm{C}}$ (\kmsi) & 4&.&689 & & &\\
$rms_{\rm{TD}}$ (d) & 0&.&0012 & & &\\
\hline
\end{tabular}
\end{minipage}
\end{table}

\subsection{The triple system KIC~5219533}\label{cas:KIC5219533}

\begin{figure*}[ht]
\begin{subfigure}[b]{0.48\textwidth}
 	\centering
  \includegraphics[width=0.89\linewidth]{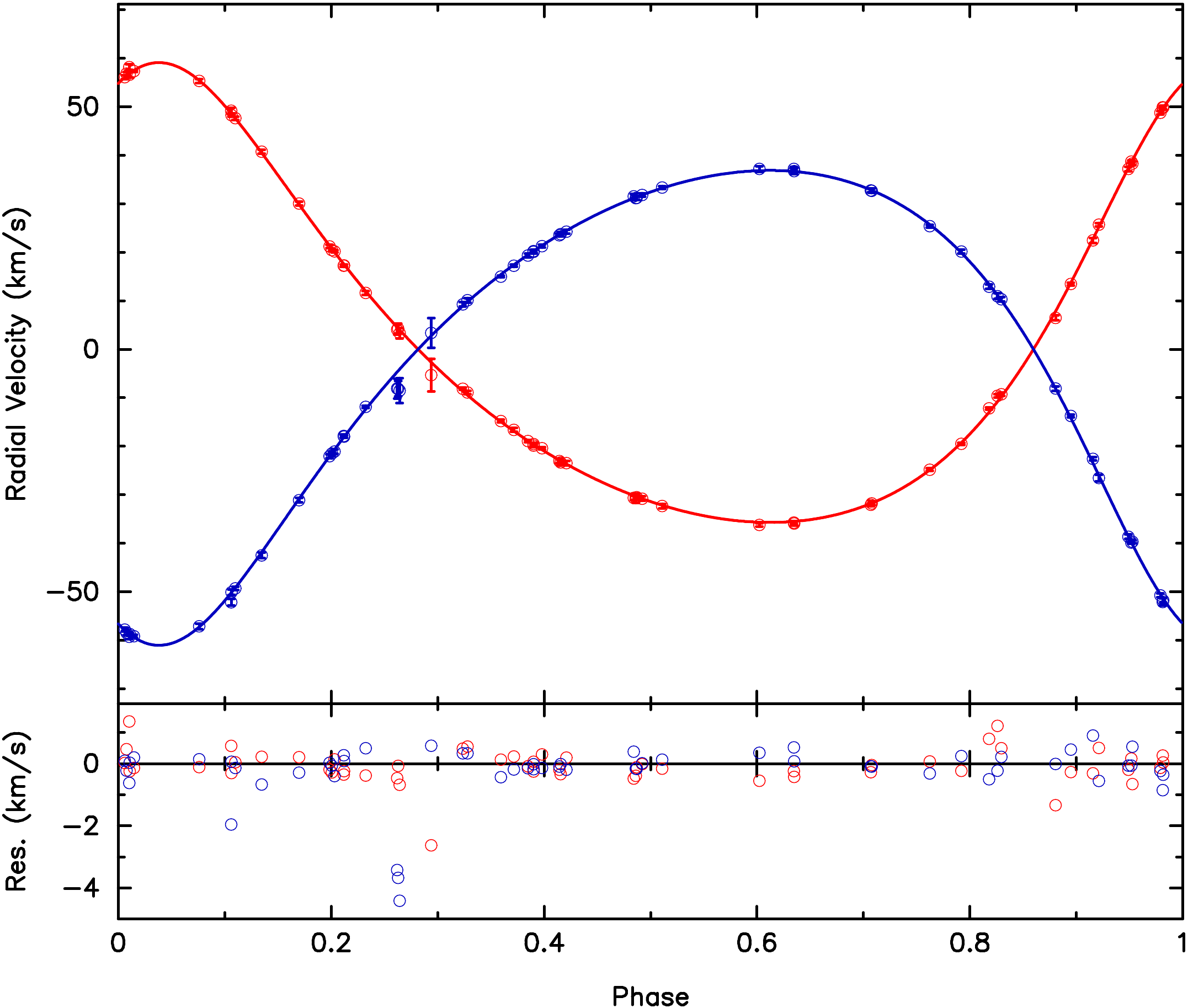}
	\label{fig:fig_KIC521a}
\end{subfigure}
\hspace{0.02\textwidth}
\begin{subfigure}[b]{0.48\textwidth}
	\centering
  \includegraphics[width=0.99\linewidth]{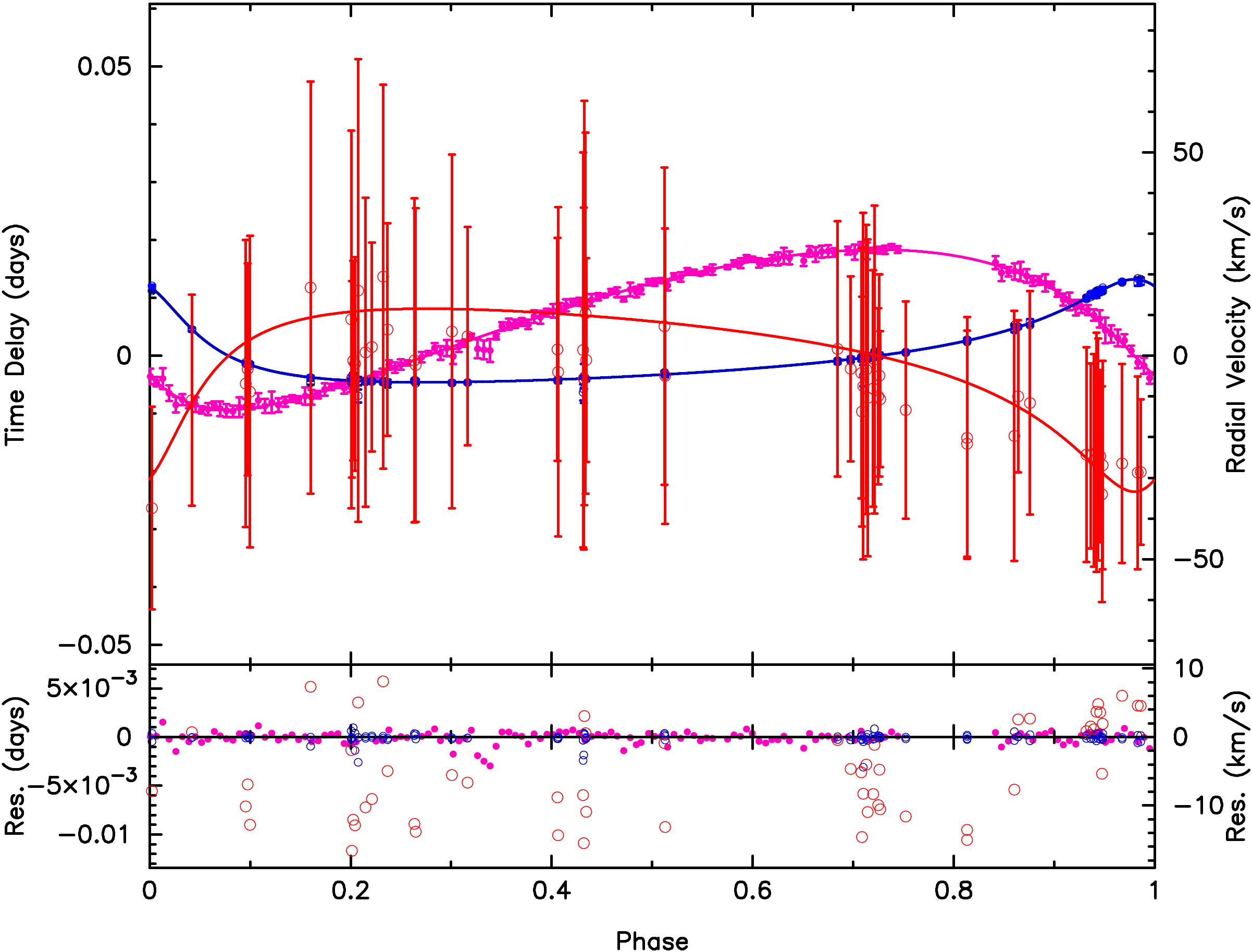}
	\label{fig:fig_KIC521b} 
\end{subfigure}
	\caption{Data and orbital solutions for KIC~5219533. {\it Left:} Inner orbit based on the HERMES RVs. The {RVs} 
	(red symbols for component A; blue symbols for component B) are overlaid with the A-B orbital solution (solid lines). 
  {\it Right:} Outer orbit based on the HERMES RVs and the {TDs} from the {\it Kepler} photometry. The {TDs and RVs} (resp. filled pink and unfilled 
  ed symbols for component C; blue symbols for the centre of mass of components A and B) are overlaid with the AB-C orbital solution (solid lines). 
	The residuals are shown in the bottom panels. }
\label{fig:fig_KIC521}
\end{figure*}

KIC~5219533 was classified as triple (SB3) with an inner pair of Am-like stars and with orbital periods of 31.919 and $\sim$ 1615 d (Sect.~\ref{KIC5219533}). 
For this system, we also 
detected long-term variations of the TDs in agreement with the long-term change observed in the RVs (Paper~I, fig.~24). We performed a simultaneous 
least-squares fitting of the updated component RVs and the mean TDs with their respective uncertainties, and obtained the best fit in the sense of min $\chi^2$ 
with the TDs assigned to the outer companion (component C). Since this component is heavily diluted in the composite spectrum, its RVs are not well-determined. 
However, a modelling of the combined data allows to derive full orbital solutions. The new parameters of the inner and outer orbital solutions are 
presented in Table~\ref{tab:KIC521}. Fig.~\ref{fig:fig_KIC521} illustrates the quality of the fit as well as both solutions. The mean residuals of the component 
RVs stay below 1~\kmsi. {The largest residuals are found near the nodal passage (both components have similar RVs)}. The orbital solution of the A-B pair 
is in good agreement with the one proposed by \citet{Catanzaro2019MNRAS.487..919C}, though our semi-major axes are slightly larger (and our component masses 
a bit smaller) than theirs. Indeed, their Figs.~3 and~4 show that several HERMES RVs (open symbols) 
lie a bit off from their final solution. The overall gain in accuracy for most parameters of the inner orbital solution is a factor of 5, while the orbital 
period is 10 times more accurate. We cannot compare the mean RV residuals since these were not displayed. In Fig.~\ref{fig:fig_KIC521} (right), the TDs and 
the RVs are plotted together with the revised AB-C orbital solution. We thus determined the semi-axis major of component C and the minimum mass of the A-B 
pair for the first time. The gain is largest for the outer orbit since the accuracy on the parameters improved with a mean factor of 10 while the orbital 
period is about 30 times more accurate compared to \citet{Catanzaro2019MNRAS.487..919C}. 
In particular, we obtained the minimum component masses $M_{\rm{A}}\,\mathrm{sin}^{3}\,i$ = 1.336 $\pm$ 0.005, $M_{\rm{B}}\,\mathrm{sin}^{3}\,i$ 
= 1.293 $\pm$ 0.004 and $M_{\rm{C}}\,\mathrm{sin}^3\,i_{\mathrm{out}}$ = 1.47 $\pm$ 0.02 M$_{\odot}$. 
From the minimum total mass of the A-B system derived for each orbit leading to the condition sin\,$i_{\mathrm{out}}$ = 0.996 ($\pm$~0.009) sin\,$i$, we 
infer that both orbits are very close to coplanarity. In summary, KIC~5219533 is a hierarchical, probably coplanar system with $q_{\mathrm{in}}$ = 0.97 and 
$q_{\mathrm{out}}$ = 0.57. We furthermore established that the $\delta$ Scuti pulsations are linked to the more massive, faster rotating star in the system. \\ 

\begin{table}
\center
\begin{minipage}{8.9cm}
\centering \caption[]{\label{tab:KIC521} Values and standard deviations of the constrained orbital parameters for KIC~5219533.}
\begin{tabular}{lr@{}l@{}lr@{}l@{}l}
\hline
\hline
\multicolumn{7}{c}{Orbital solution A-B}\\
\hline
Orbital parameter &\multicolumn{3}{c}{Value} & \multicolumn{3}{c}{Std. dev.}\\
\hline
$P$ (d)&31&.&91763 &0&.&00022\\
$T_{\rm{0}}$ (Hel. JD)& 57467&.&41 &0&.&02\\
$e$  &0&.&271 &0&.&001\\
$\omega$ (\degr)&335&.&3 &0&.&2\\
$V_{\gamma}$ (\kmsi)& (var.)& & & & &\\
$a_{\rm{A}} \mathrm{sin}\,i$ (au) &0&.&13369 &0&.&00019\\ 
$a_{\rm{B}} \mathrm{sin}\,i$ (au) &0&.&13810 &0&.&00023\\ 
\hline
$K_{\rm{A}}$ (\kmsi) & 47&.&33 & 0&.&07\\
$K_{\rm{B}}$ (\kmsi) & 48&.&90 & 0&.&08\\
$M_{\rm{A}} \mathrm{sin}^3\,i$ (M$_\odot$) &1&.&336 &0&.&005\\
$M_{\rm{B}} \mathrm{sin}^3\,i$ (M$_\odot$) &1&.&293 &0&.&004\\
$q_{\rm{in}}$ &  0&.&968 & 0&.&002 \\
\hline
$rms_{\rm{A}}$ (\kmsi) & 0&.&550 & & &\\
$rms_{\rm{B}}$ (\kmsi) & 0&.&833 & & &\\
\hline
\hline
\multicolumn{7}{c}{Orbital solution AB-C}\\
\hline
Orbital parameter &\multicolumn{3}{c}{Value} & \multicolumn{3}{c}{Std. dev.}\\
\hline
$P$ (d)&1603&.&5 &1&.&7\\
$T_{\rm{0}}$ (Hel. JD)& 55212&.&9 &3&.&2\\
$e$  &0&.&564 &0&.&004\\
$\omega$ (\degr)&210&.&6 &0&.&5\\
$V_{\gamma}$ (\kmsi)&10&.&70 & 0&.&04\\
$a_{\rm{AB}} \mathrm{sin}\,i_{\mathrm{out}}$ (au) &1&.&542 &0&.&008\\ 
$a_{\rm{C}} \mathrm{sin}\,i_{\mathrm{out}}$ (au) &2&.&74 &0&.&03\\ 
\hline
$a_{\rm{TD}_{\rm{C}}}/c\, \mathrm{sin}\,i_{\mathrm{out}}$ (d) & 0&.&0158 & 0&.&0002\\
$K_{\rm{AB}}$ (\kmsi) & 12&.&67 & 0&.&08\\
$K_{\rm{C}}$ (\kmsi) & 22&.&49 & 0&.&27\\
$M_{\rm{AB}} \mathrm{sin}^3\,i_{\mathrm{out}}$ (M$_\odot$) &2&.&60 &0&.&07\\
$M_{\rm{C}} \mathrm{sin}^3\,i_{\mathrm{out}}$ (M$_\odot$) &1&.&47 &0&.&02\\
$q_{\rm{out}}$ &  0&.&565 & 0&.&006 \\
\hline
$rms_{\rm{C}}$ (\kmsi) & 8&.&294 & & &\\
$rms_{\rm{TD}}$ (d) & 0&.&0006 & & &\\
\hline
\end{tabular}
\end{minipage}
\end{table}

\subsection{KIC~8975515}\label{cas:KIC8975515}

\begin{figure*}[ht]
\begin{subfigure}[b]{0.48\textwidth}
	\centering
  \includegraphics[width=0.89\linewidth]{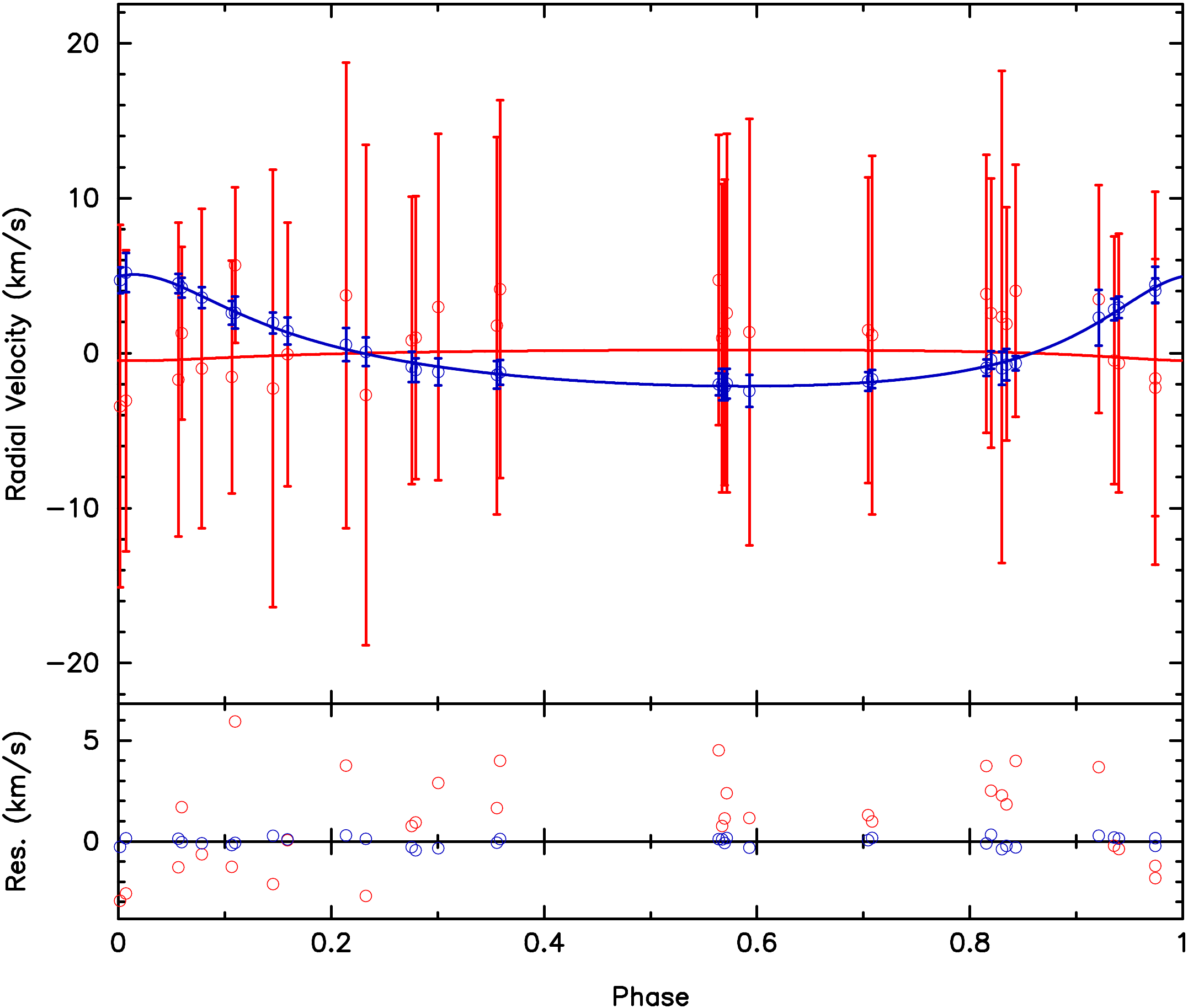}
	\label{fig:fig_KIC897o}
\end{subfigure}
\hspace{0.02\textwidth}
\begin{subfigure}[b]{0.48\textwidth}
 	\centering
  \includegraphics[width=0.99\linewidth]{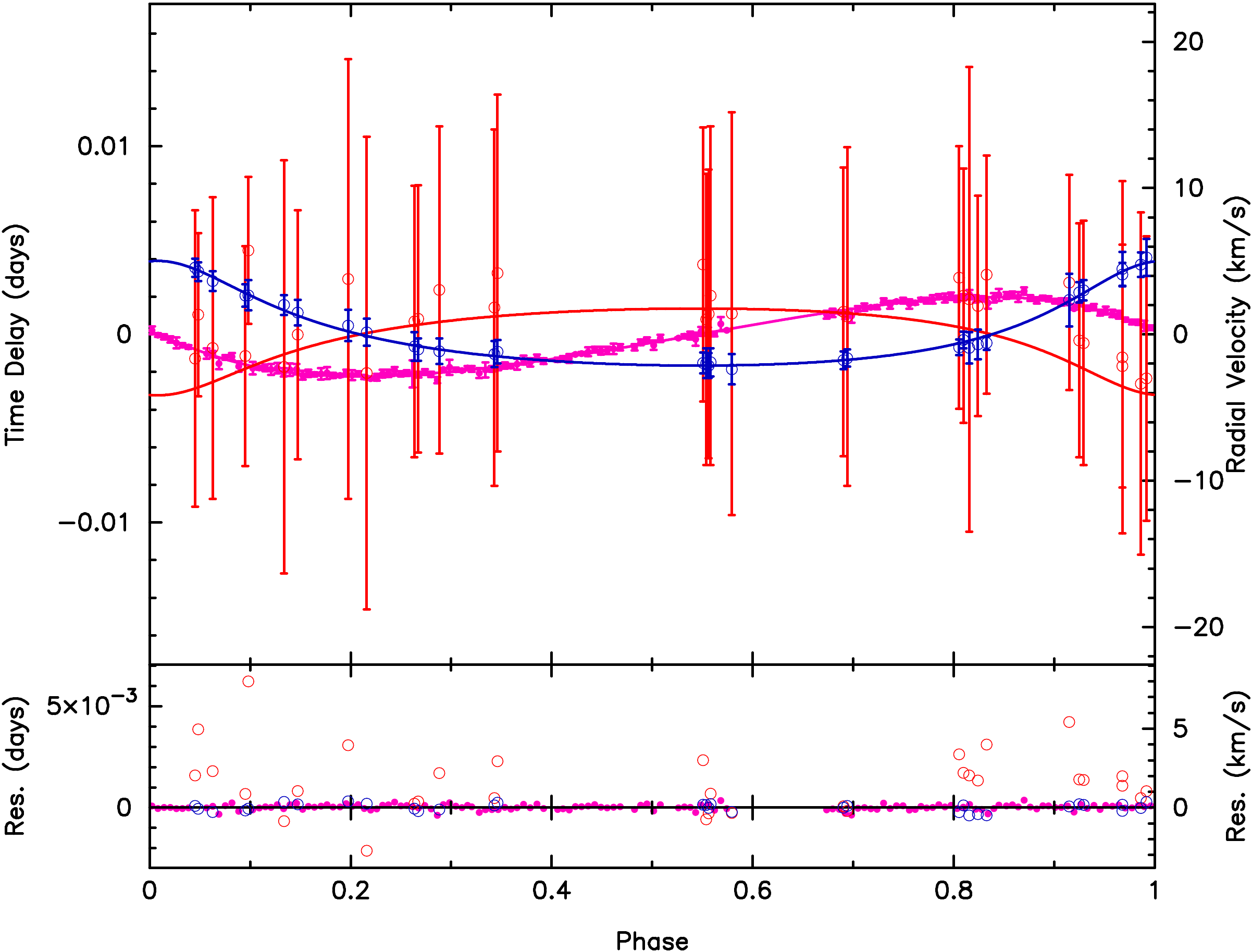}
	\label{fig:fig_KIC897n}
\end{subfigure}
	\caption{Data and orbital solution for KIC~8975515. {\it Left:} Orbit (nr 1) based on the HERMES RVs. The RVs (red symbols for component A; 
	blue symbols for component B) are overlaid with a preliminary orbital solution of type SB2, however without much constraint on the amplitude 
	K$_{A}$. {\it Right:} Orbit (nr 2) based on the HERMES RVs and the TDs from the {\it Kepler} photometry. The {TDs and RVs} 
	(resp. filled pink and unfilled red symbols for component A; blue symbols for component B) are overlaid with the A-B orbital solution (solid lines). 
	The residuals are shown in the bottom panels.}
	\label{fig:fig_KIC897}
\end{figure*}

This star was recognized as SB2 from our multi-epoch study, and also shows long-term variations of the TDs coherent with the variations found 
in the RVs (Paper~I, fig.~24). The system consists of late A-type stars with one fast spinning (component A) and one (apparently) 
slowly spinning star (component B). In Paper~I, we derived an orbital solution of type SB1 based on the RVs of component B only (\porb
$\sim$ 1600 d). 
We first derived a pure RV-based orbital solution (using the RVs of both components). Unfortunately, the extreme uncertainties on the RVs of 
the fast spinning component {did not} allow to constrain its RV amplitude. This is obvious in Fig.~\ref{fig:fig_KIC897} (left) which illustrates 
the best-fit orbital solution with a period of 1581 d. Subsequently, we performed two simultaneous (RV+TD) analyses, by associating the TDs 
with each set of RVs in turn, and found an optimum fit when the TDs were assigned to component A. 

We determined an orbital period of 1603.4 $\pm$ 9.3 d. The new orbital parameters and their derived fundamental properties are 
listed in Table~\ref{tab:KIC8975515}. The residuals show the very good agreement between both data types, with small mean residuals for the 
TDs and RVs of component B (the slower rotator), and a large (non-zero) mean residual for component A (the faster rotator). 
The orbital solution is illustrated in Fig.~\ref{fig:fig_KIC897} (right). In this case, the mass ratio ($q$ = 0.83 $\pm$ 0.05) is well-determined 
thanks to the high accuracy of the TDs (whereas the corresponding component RVs show a very large scatter). The present method allowed us to derive 
information which was otherwise impossible to obtain with a pure RV-based solution. Another conclusion is that the faster rotating component 
exhibits the $\delta$ Scuti pulsations. 
This may also explain why the RV residuals of component A are found to be systematically off from the final solution.
From the parameters displayed in Table~\ref{tab:KIC8975515}, we obtain the minimum masses, $M_{\rm{A}}\,\mathrm{sin}^3\,i$ = 0.0195 
$\pm$ 0.0024 and $M_{\rm{B}}\,\mathrm{sin}^3\,i$ = 0.0161 $\pm$ 0.0011 M$_{\odot}$. Assuming that either one of the components has a typical 
mass of $\sim$ 2 M$_{\odot}$, we expect that the system is seen under a low inclination angle ($i$ $\sim$12\degr).\\

\begin{table}
\center
\begin{minipage}{8.9cm}
\centering \caption[]{\label{tab:KIC8975515} Values and standard deviations of the constrained orbital parameters for KIC~8975515.}
\begin{tabular}{lr@{}l@{}lr@{}l@{}l}
\hline
\hline
\multicolumn{7}{c}{Orbital solution A-B}\\
\hline
Orbital parameter &\multicolumn{3}{c}{Value} & \multicolumn{3}{c}{Std. dev.}\\
\hline
$P$ (d)&1603&.&4 &9&.&3\\
$T_{\rm{0}}$ (Hel. JD)& 58690&.& &19&.&\\
$e$  &0&.&408 &0&.&015\\
$\omega$ (\degr)& 172&.&5 &2&.&6\\
$V_{\gamma}$ (\kmsi)&-20&.&63 & 0&.&14\\
$a_{\rm{A}} \mathrm{sin}\,i$ (au) &0&.&3988 &0&.&0053\\ 
$a_{\rm{B}} \mathrm{sin}\,i$ (au) &0&.&483 &0&.&028\\ 
\hline
$a_{\rm{TD}_{\rm{A}}}/c\, \mathrm{sin}\,i$(d) & 0&.&00230 & 0&.&00003\\
$K_{\rm{A}}$ (\kmsi) & 2&.&96 & 0&.&05\\
$K_{\rm{B}}$ (\kmsi) & 3&.&59 & 0&.&21\\
$M_{\rm{A}} \mathrm{sin}^3\,i$ (M$_\odot$) &0&.&0195 &0&.&0024\\
$M_{\rm{B}} \mathrm{sin}^3\,i$ (M$_\odot$) &0&.&0161 &0&.&0011\\
$q$ & 0&.&83 & 0&.&05 \\
\hline
$rms_{\rm{A}}$ (\kmsi) & 2&.&597 & & &\\
$rms_{\rm{B}}$ (\kmsi) & 0&.&237 & & &\\
$rms_{\rm{TD}}$ (days) & 0&.&00013 & & &\\
\hline
\end{tabular}
\end{minipage}
\end{table}
 
\subsection{KIC~9775454}\label{cas:KIC9775454}

\begin{figure}[ht]
 	\centering
	\includegraphics[width=0.95\linewidth]{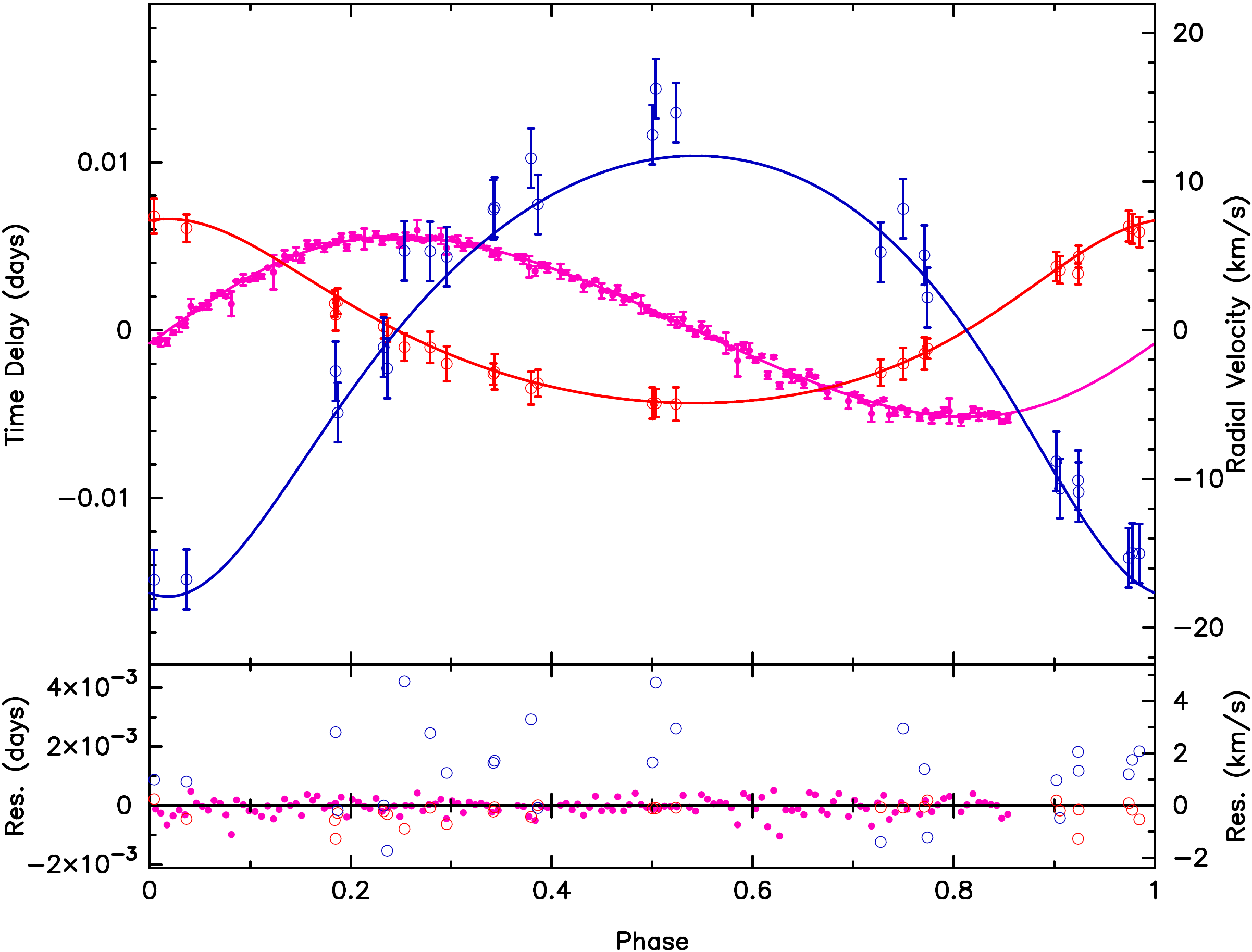} 
	\caption{Data and orbital solution for KIC~9775454. The {TDs} (filled pink symbols) and the {RVs} (unfilled red symbols for 
  component A; blue symbols for component B) are overlaid with the A-B orbital solution (solid lines).
	The residuals are shown in the bottom panel. }
	\label{fig:fig_KIC977}
\end{figure}

\begin{figure}[ht]
	\centering
	\includegraphics[width=0.9\linewidth]{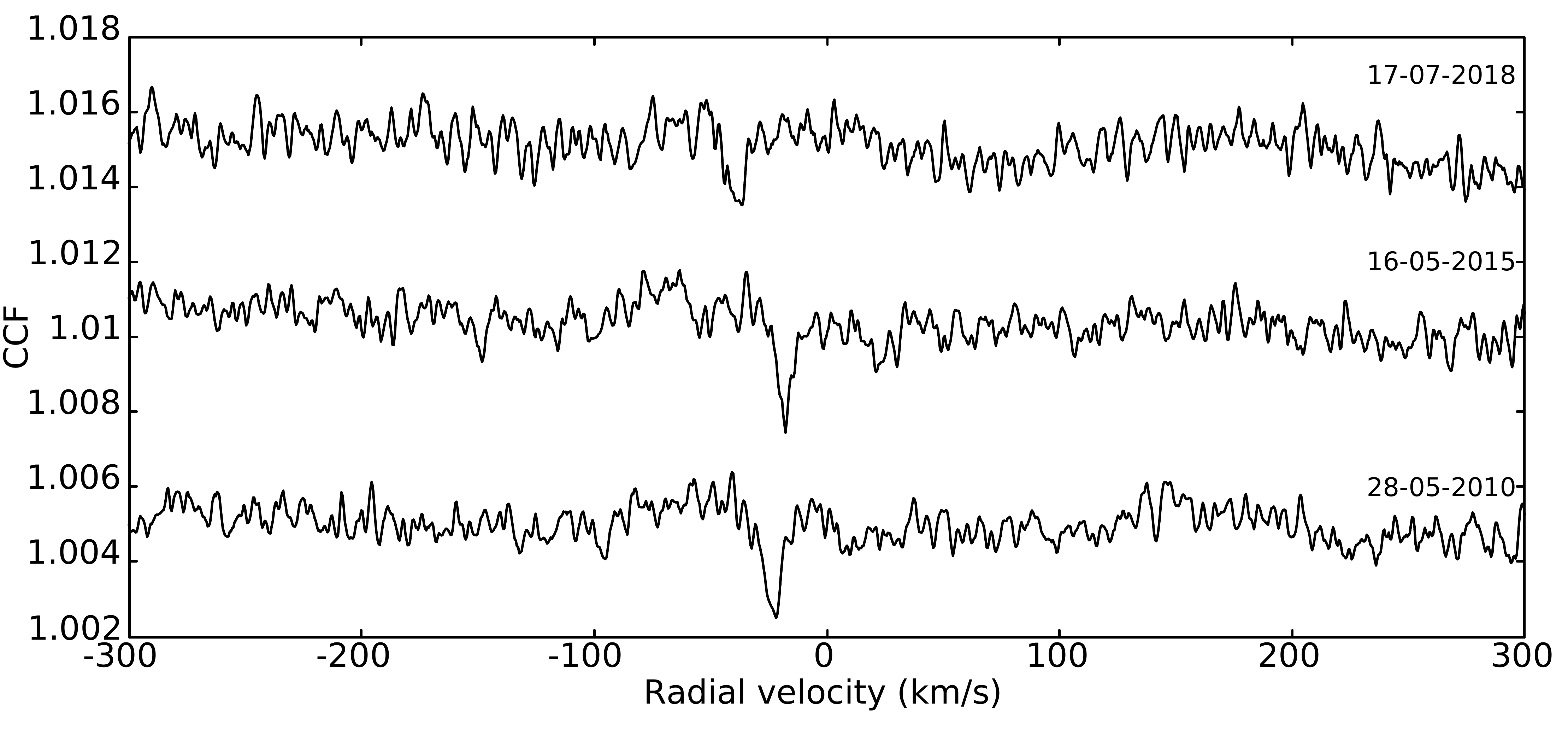}
	\caption{CCFs of the residual spectra of KIC~9775454 computed with a mask of type K0 for three different dates. } 
 	\label{fig:ccf_KIC977}
\end{figure}

KIC~9775454 was classified as a long-period SB1 (Sect.~\ref{KIC9775454}). It was shown to display long-term variations of the TDs in agreement 
with the orbital variations of the RVs (this concerns also the dominant frequency located at 4.161 d$^{-1}$, cf. Paper~I). \citet{Murphy2018MNRAS.474.4322M}
used the HERMES RVs of the broad-lined primary (component~A) and derived an orbital solution with a period of $\sim$1700 days from their combined (RV+TD)
analysis. Being much fainter, component~B was not directly detected in the high-resolution spectra. Nevertheless, we managed to {find an indication 
of a cool companion by computing CCFs of the residual spectra (after proper subtraction of an adequate synthetic spectrum 
to remove the contribution of the primary) using a mask of spectral type K0. } The ``residual''  CCFs were computed for the spectral range [5100 - 5700] \AA,
and revealed {a shallow peak that reflects the contribution of the companion and displays Doppler shifts} (Fig.~\ref{fig:ccf_KIC977}). 
In this way, we were able to extract additional RV data which behave in anti-{phase} with those of component~A {(Fig.~\ref{fig:fig_KIC977})}. We roughly 
estimated their uncertainties from the widths of the associated spectral lines. Thus, we can classify KIC~9775454 as an SB2. We next performed a simultaneous 
(RV+TD) analysis, by fitting the component RVs together with the TDs assigned to the broad-lined primary (component~A). The new orbital solution is 
illustrated by Fig.~\ref{fig:fig_KIC977}. The RV residuals of component~A and the TD residuals are within expectation while the RV residuals of component~B 
are systematically large. This can be explained by the noisy ``residual'' CCFs associated to the difficulty of identifying the lines of the cool component 
in the spectra. The orbital parameters and derived fundamental properties are listed in Table~\ref{tab:KIC9775454}. These parameters agree very well 
with the values proposed by \citet{Murphy2018MNRAS.474.4322M}. By treating this system as an SB2 and using the TDs, we determined for the first time a 
reasonable value of the mass ratio ($q$ = 0.42 $\pm$ 0.02). In this special case where the TDs are linked with both the low- and the high-frequency regions, 
the more massive primary component is very likely the hybrid pulsator. \\ 

\begin{table}
\center
\begin{minipage}{8.9cm}
\centering \caption[]{\label{tab:KIC9775454} Values and standard deviations of the constrained parameters of the orbital 
solution for KIC~9775454.}
\begin{tabular}{lr@{}l@{}lr@{}l@{}l}
\hline
\hline
\multicolumn{7}{c}{Orbital solution A-B}\\
\hline
Orbital parameter &\multicolumn{3}{c}{Value} & \multicolumn{3}{c}{Std. dev.}\\
\hline
$P$ (d)&1706&.&8 &6&.&4\\
$T_{\rm{0}}$ (Hel. JD)& 56655&.& &10&.&\\
$e$  &0&.&212 &0&.&008\\
$\omega$ (\degr)& 349&.&6 &1&.&9\\
$V_{\gamma}$ (\kmsi)&-22&.&15 & 0&.&16\\
$a_{\rm{A}} \mathrm{sin}\,i$ (au) &0&.&949 &0&.&017\\ 
$a_{\rm{B}} \mathrm{sin}\,i$ (au) &2&.&27 &0&.&09\\ 
\hline
$a_{\rm{TD}_{\rm{A}}}/c\, \mathrm{sin}\,i$ (d) & 0&.&00548 & 0&.&00010\\
$K_{\rm{A}}$ (\kmsi) &  6&.&19 & 0&.&12\\
$K_{\rm{B}}$ (\kmsi) & 14&.&82 & 0&.&62\\
$M_{\rm{A}} \mathrm{sin}^3\,i$ (M$_\odot$) &1&.&08 &0&.&11\\
$M_{\rm{B}} \mathrm{sin}^3\,i$ (M$_\odot$) &0&.&45 &0&.&03\\
$q$ & 0&.&42 &  0&.&02 \\
\hline
$rms_{\rm{A}}$ (\kmsi) & 0&.&471 & & &\\
$rms_{\rm{B}}$ (\kmsi) & 2&.&150 & & &\\
$rms_{\rm{TD}}$ (d) & 0&.&00028 & & &\\
\hline
\end{tabular}
\end{minipage}
\end{table}

\section{Search for regular frequency patterns}\label{sec:frequent_spacings}

In this section, we investigate the distributions of the frequency differences (aka spacings) that occur in the high-frequency regime 
of the Fourier transforms of each system. The aim is to identify the {(most)} frequently occurring frequency spacings for each object, 
and to verify whether some {of these} spacings might be related to the orbital or the rotational periods. In the case of an orbital or 
a rotational origin of the low frequencies (e.g. due to modulations in the light curve caused by the tidal deformations or by the presence 
of spots or other surface inhomogeneities e.g. MOBSTER \citep{2019MNRAS.487.4695S} or caused by {the mechanism of} tidal excitation 
e.g. KIC~4142768 \citep{2019ApJ...885...46G}), we may expect to detect a significant number of harmonics of the corresponding main 
frequencies. In the case of a tidal or rotational splitting of the high frequencies {associated to the} rapid pulsations \citep[e.g. 
of tidal splitting {of $p$-modes} in KIC~6048106 and U~Gru, respectively,][]{2018AcA....68..425S, Bowman2019ApJ...883L..26B}, 
we may expect to find regular {(also non-harmonic)} frequency patterns whose spacing values {will point} at those frequencies. \\

\begin{figure*}[ht]
	\centering
 \begin{subfigure}[b]{0.48\textwidth}
  \centering
  \includegraphics[width=1.0\linewidth]{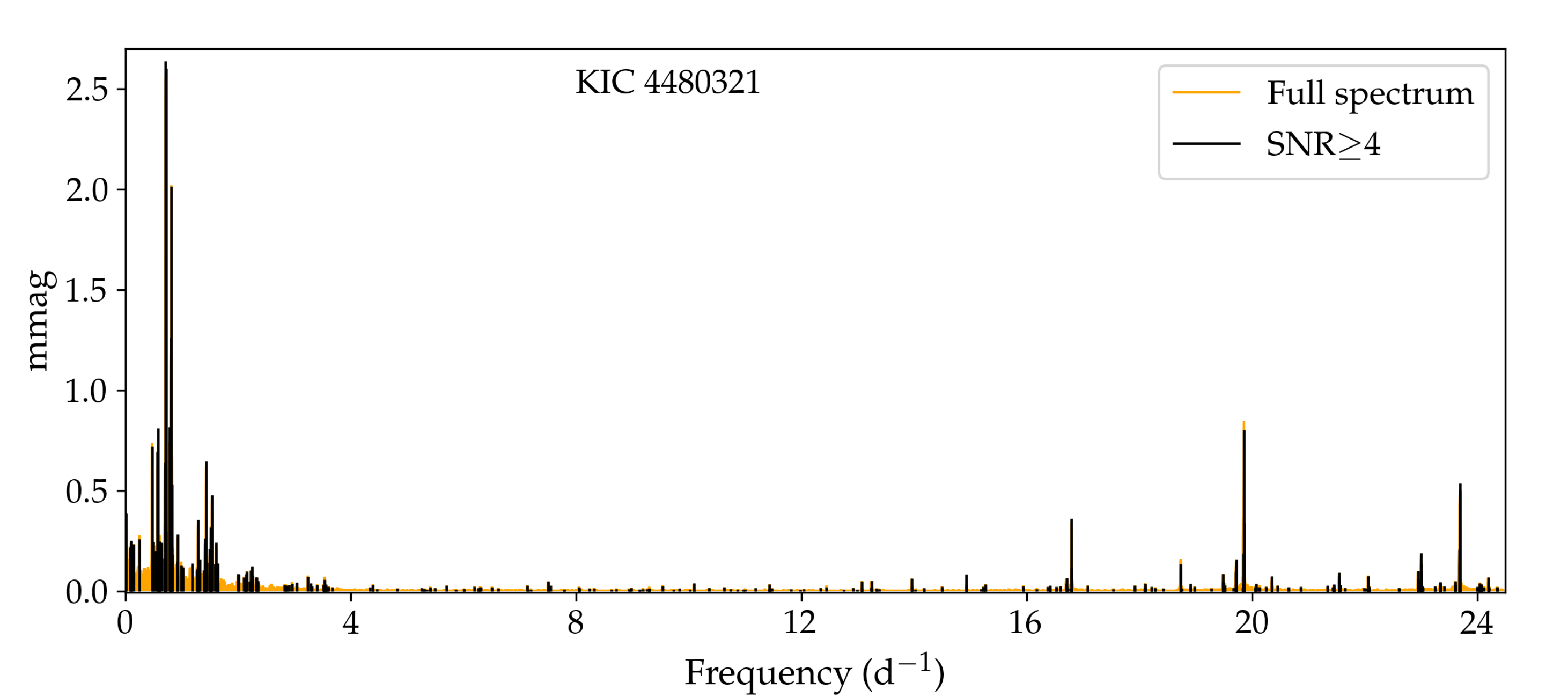}
 	\label{fig:perK448}
 \end{subfigure}
 \hspace{0.02\textwidth} 
 \begin{subfigure}[b]{0.48\textwidth}
  \centering
  \includegraphics[width=1.0\linewidth]{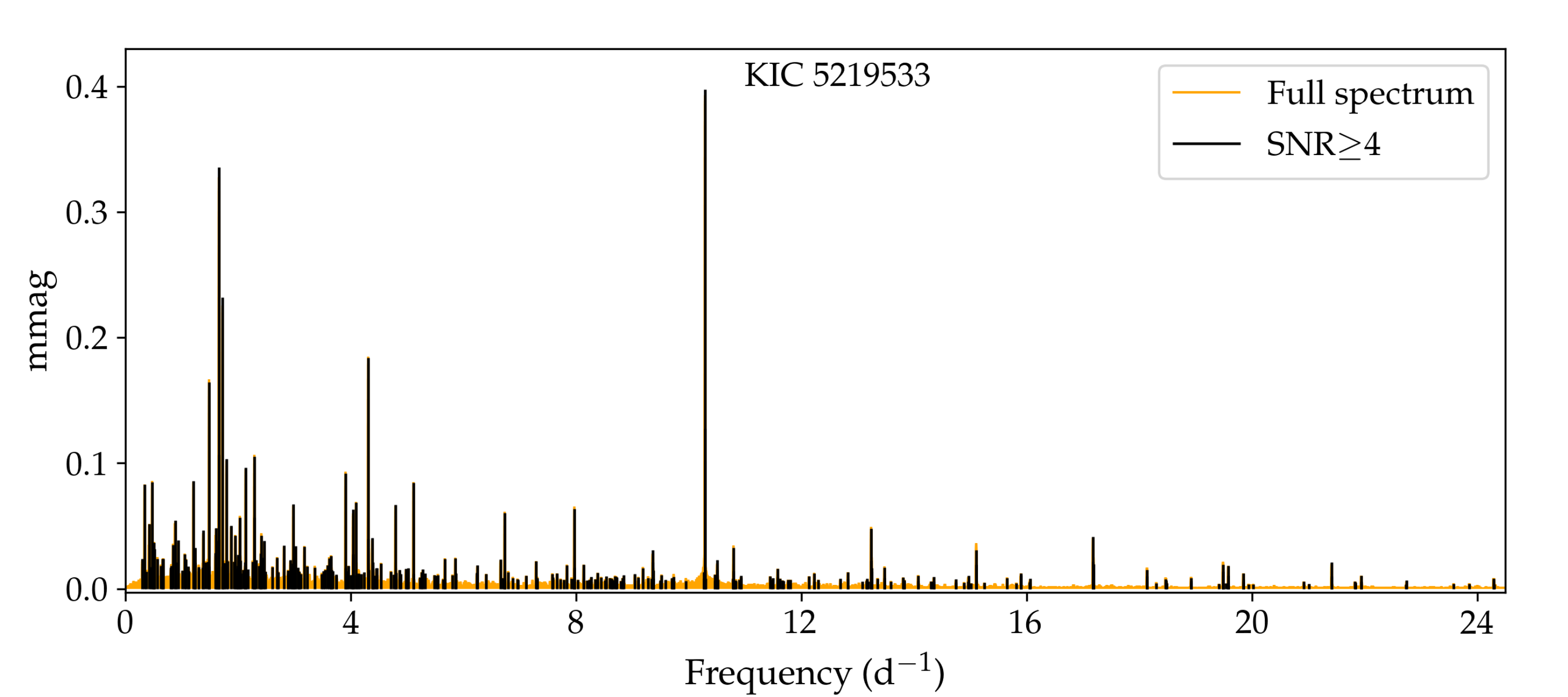}
 	\label{fig:perK521}
 \end{subfigure}
 \begin{subfigure}[b]{0.48\textwidth}
 \centering
  \includegraphics[width=1.0\linewidth]{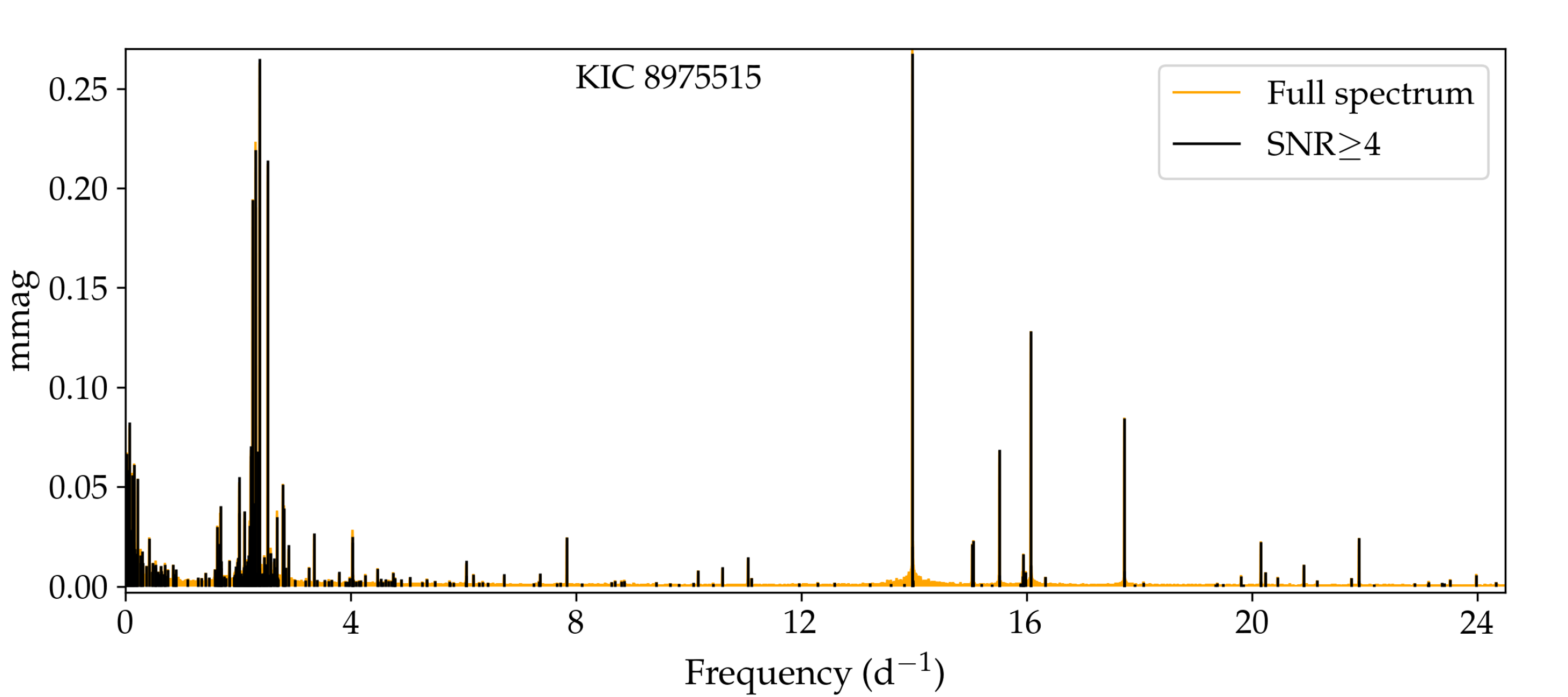}
 	\label{fig:perK897}
 \end{subfigure}
 \hspace{0.02\textwidth} 
 \begin{subfigure}[b]{0.48\textwidth}
 \centering
	\includegraphics[width=1.0\linewidth]{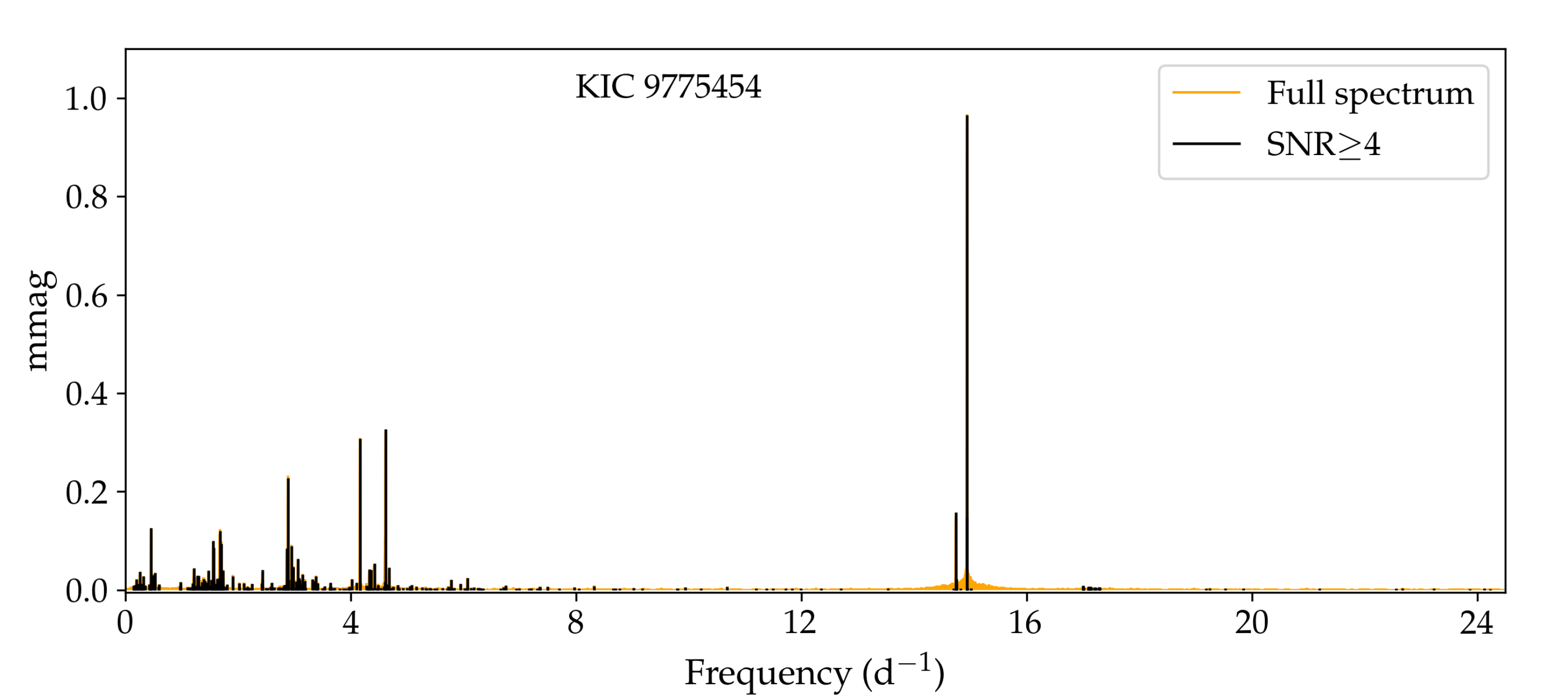}
 	\label{fig:perK977}
 \end{subfigure}
	\caption{Lomb-Scargle periodograms derived from the {\it Kepler} data for the systems {KIC~4480321 (top left),~5219533 (top right),
  ~8975515 (bottom left) and KIC~9775454 (bottom right)}. The frequencies plotted in orange illustrate the full spectrum {across the range 
  [0 - 24.5]~d$^{-1}$}, while {the frequencies overplotted in black have} a signal-to-noise ratio higher than 4. }
	\label{fig:period-grams}
\end{figure*}

We used the detrending algorithm of Lightkurve v1.9 to detrend the {\it Kepler} light curves of the previously discussed systems 
\citep{2020AAS...23540904B}, and computed the periodograms with the Lomb-Scargle technique (Fig.~\ref{fig:period-grams}). 
{For each system,} we analysed {the frequency interval of the periodogram where most of the power is located}. For KIC~4480321, 
this corresponds to the [12.5, 24.7]~d$^{-1}$ interval; for KIC~5219533, this is the [6, 22.5]~d$^{-1}$ interval; for KIC~8975515, the 
[12.5, 17.5]~d$^{-1}$ interval, and for KIC~9775454, the [5.7, 22.5]~d$^{-1}$ interval. We then computed the spacings between {all 
possible pairs of significant frequencies} while {considering} the limiting {frequency} resolution of the {\it Kepler} long-cadence 
data sets, and counted the number of {individual} occurrences \citep[as in][]{2018AcA....68..425S}. 

\begin{figure*}[ht]
	\centering
 \begin{subfigure}[b]{0.48\textwidth}
  \centering
  \includegraphics[width=1.0\linewidth]{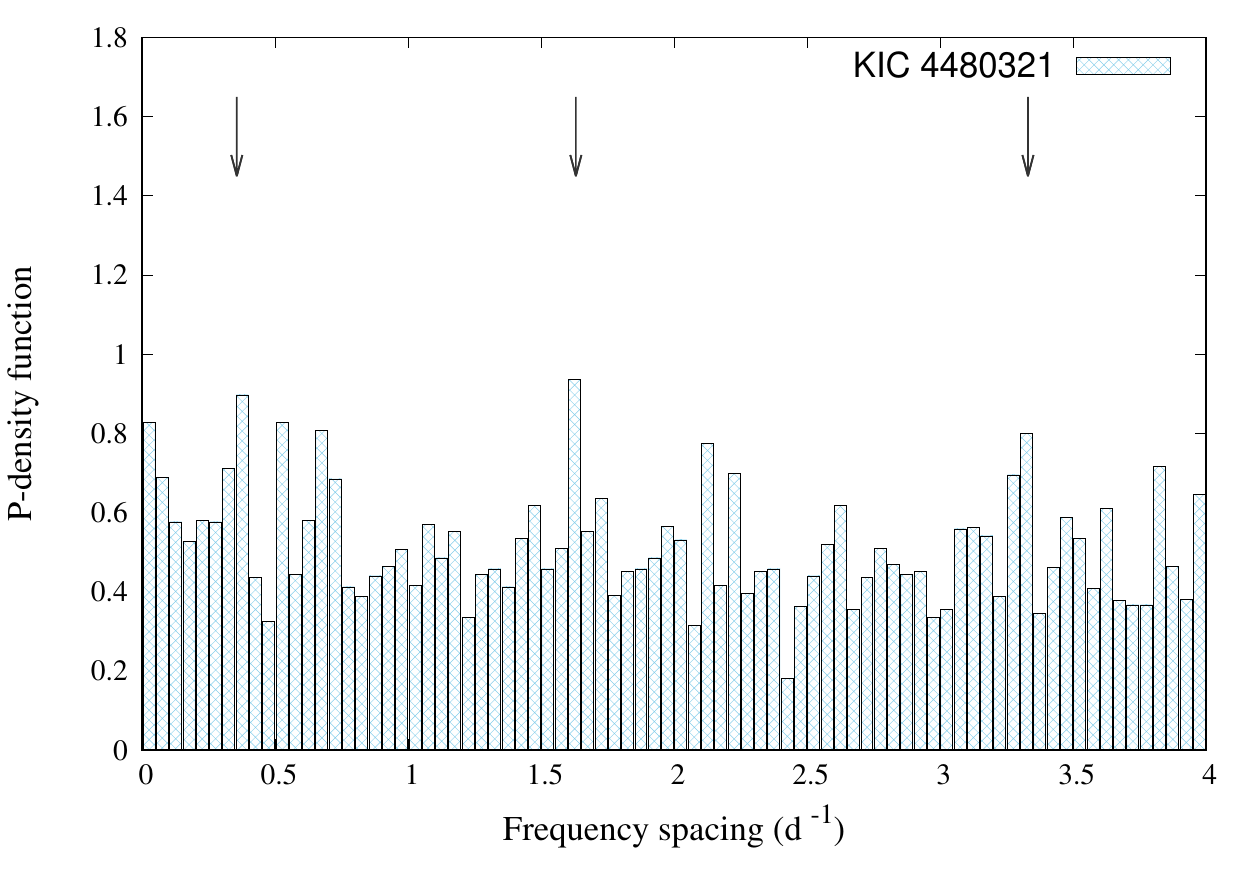}
 	\label{fig:histoK448}
 \end{subfigure}
 \hspace{0.02\textwidth} 
 \begin{subfigure}[b]{0.48\textwidth}
 \centering
	\includegraphics[width=1.0\linewidth]{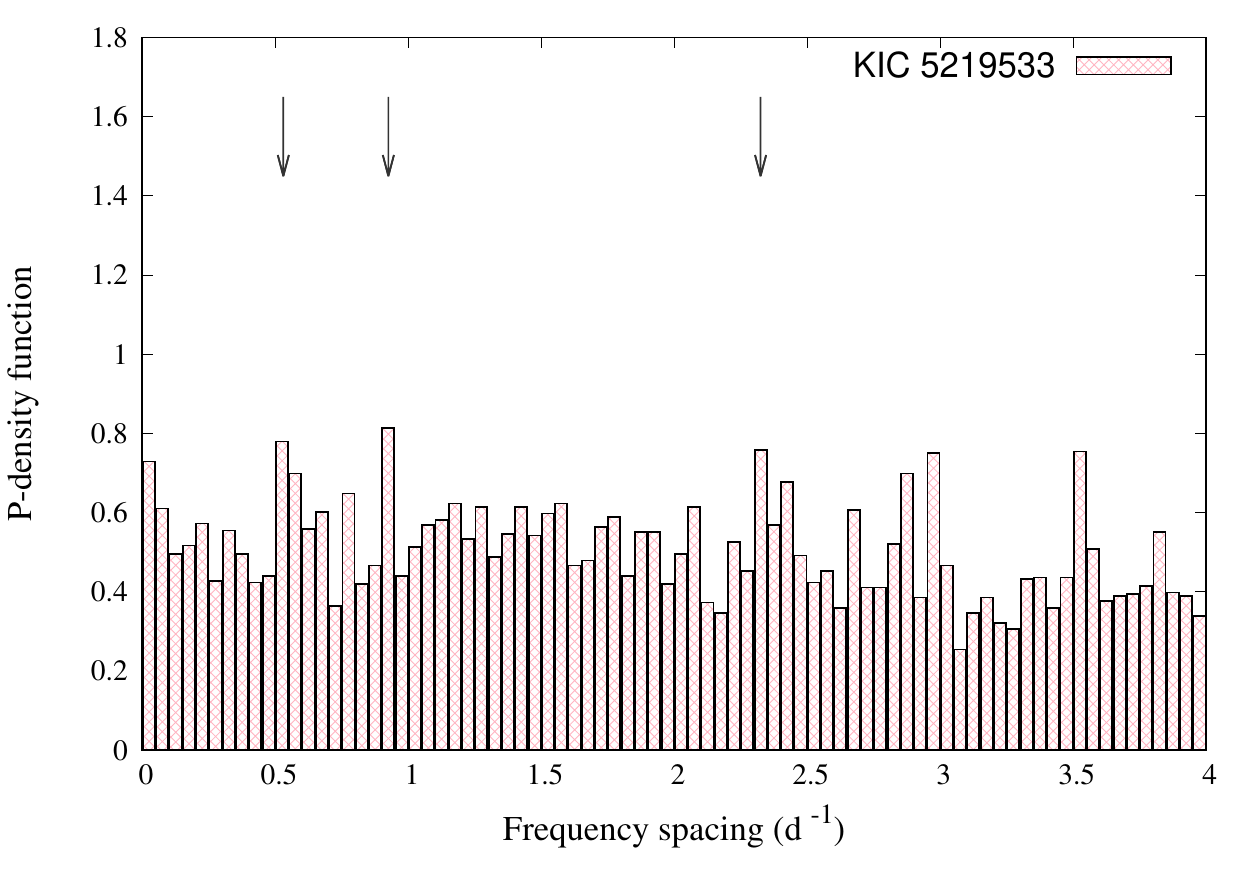}
 	\label{fig:histoK521}
 \end{subfigure}
 \begin{subfigure}[b]{0.48\textwidth}
 \centering
	\includegraphics[width=1.0\linewidth]{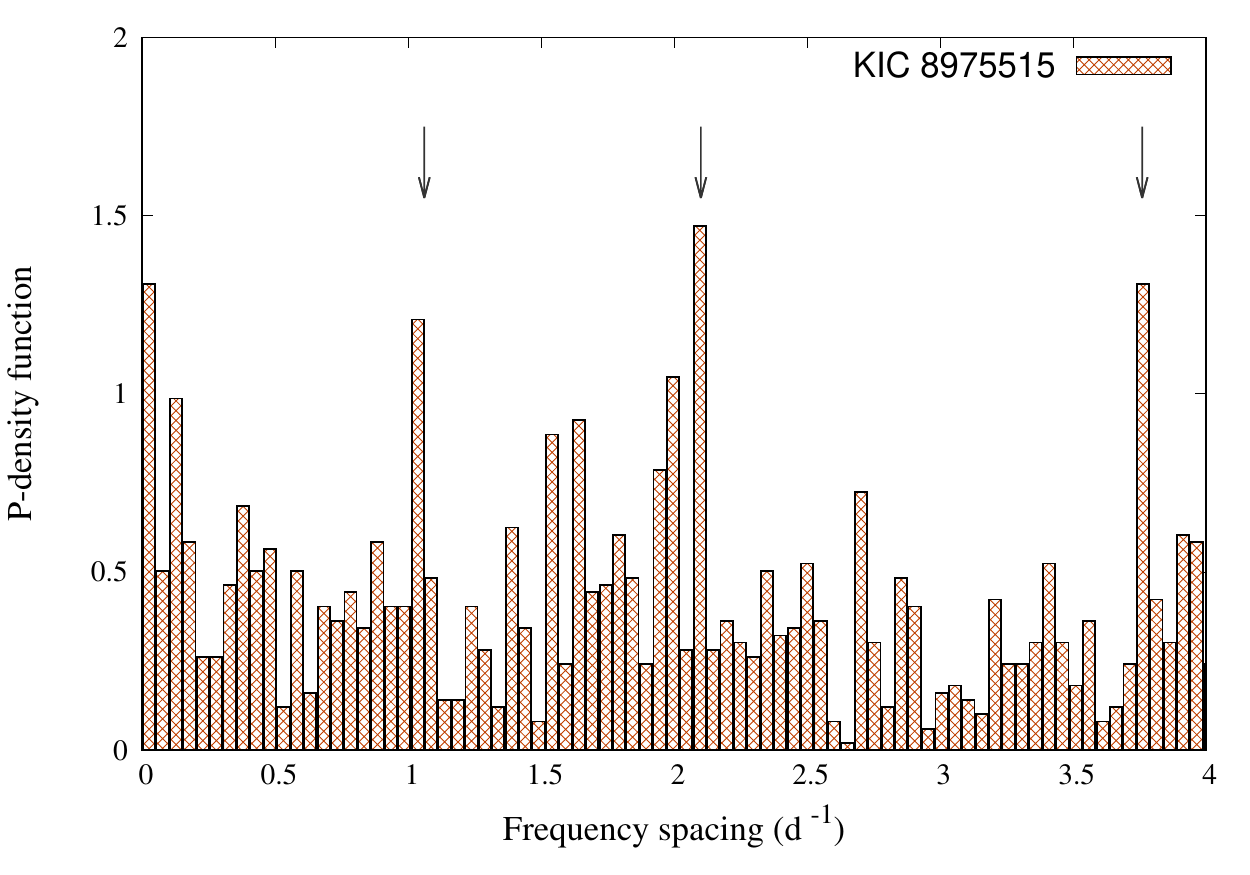}
 	\label{fig:histoK897}
 \end{subfigure}
 \hspace{0.02\textwidth} 
 \begin{subfigure}[b]{0.48\textwidth}
 \centering
	\includegraphics[width=1.0\linewidth]{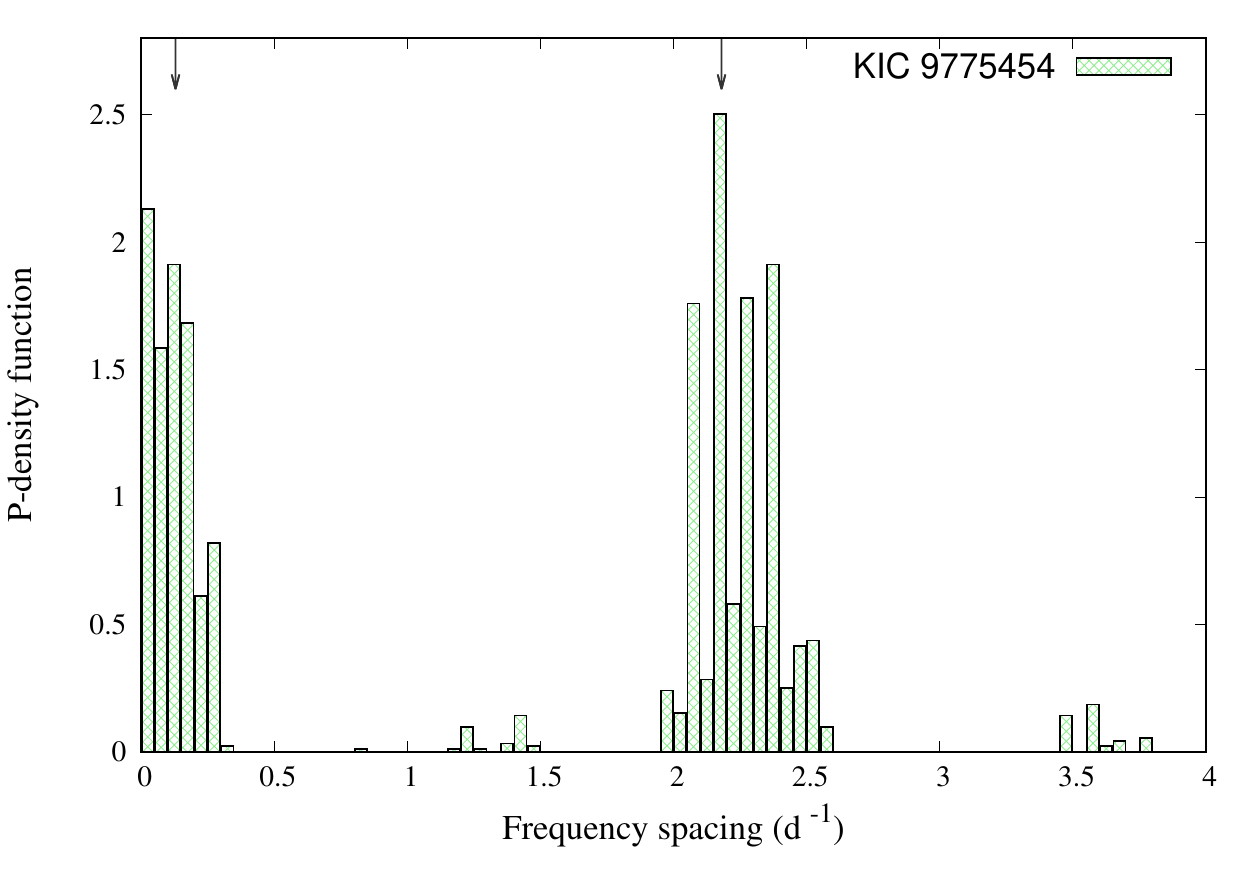}
 	\label{fig:histoK977}
 \end{subfigure}
	\caption{Normalized density functions of the frequency spacings {across the range [0 - 4]~d$^{-1}$} derived from the 
  {\it Kepler}-based periodograms for the systems {KIC~4480321 (top left),~5219533 (top right),~8975515 (bottom left) 
  and KIC~9775454 (bottom right)}. The highest occurrences are marked by grey arrows. } 
	\label{fig:histo-spacings}
\end{figure*}

We next discuss the {density distributions} of the frequency spacings derived from the {\it Kepler}-based periodograms 
for the four systems with the improved orbits. The density plots of the frequency spacings were produced from the counts 
adopting a bin size of 0.05~d$^{-1}$ in each case. In Fig.~\ref{fig:histo-spacings}, we show the {density plots} of the 
spacings for KIC~4480321,~5219533,~8975515 and KIC~9775454 in the restricted frequency range [0 - 4]~d$^{-1}$, i.e. the interval 
where the highest occurrences can be found. A general observation is that the distributions of the frequency spacings in the 
high-frequency regime of the four systems show large differences between them. Also, {a high(er) peak in the distributions 
(or a secondary maximum) of} individual spacings is generally located in the first bin. This is due to the fact that this bin 
also contains the smallest possible spacing which corresponds to the frequency resolution of 0.00068~d$^{-1}$ {in the 
{\it Kepler} long-cadence data}. 

From Fig.~\ref{fig:histo-spacings}, we derived the following findings:\\  
- {Both KIC~4480321 and KIC~5219533 present a dense and continuous distribution of frequency spacings with several 
(almost) equal {local} maxima. There is power everywhere, also in the first bin. The close binaries of these triple 
systems have orbital frequencies of respectively 0.11~d$^{-1}$ and 0.03~d$^{-1}$, which means that in the case of KIC~4480321, 
the bin size is small enough to draw a firm conclusion with respect to tidal splitting. In the case of the remaining binary 
systems, the orbital periods are of order of several years and too large to be resolved in the periodograms.} In the case 
of KIC~4480321, the highest occurrences 
occur at the spacings of 0.35, 1.63 (highest), and 3.33~d$^{-1}$. The latter spacing shows a harmonic relation with 1.63~d$^{-1}$ 
(ratio of almost 2). {This interdependence between two frequent spacings could be very well explained by rotation.} 
{Also,} \citet{Li2020MNRAS.491.3586L} derived 0.0070 $\pm$ 0.0007~d$^{-1}$ for the near-core rotation rate from a modelling 
of the slope-period relation based on the regular period spacings in the $g$-mode region. Such a {fast rotation rate} stands in 
sharp contrast with the {the abovelisted most frequent spacings if we consider them as} potential surface rotation rates for 
component~C (the fast-spinning component). On the other hand, {we find no} peak occurring near or at the orbital frequency 
value of 0.11~d$^{-1}$. 
We conclude that the orbital frequency of the inner binary system does not affect the pulsation frequencies in the $p$-mode region. 
This is perfectly consistent with the conclusion that component~C is the $\delta$ Scuti pulsator. \\  
- In the case of KIC~5219533, the most frequent spacings occur at the frequencies of 0.55, 0.94 (highest) and 2.32~d$^{-1}$, 
{(or the equally frequent spacings at 2.98 and 3.59~d$^{-1}$}).  
The frequent spacings of 0.55 and 0.94~d$^{-1}$ could indicate the rotation period of component~C (the fast-spinning component). 
We note that \citet{Li2020MNRAS.491.3586L} derived 0.59 $\pm$ 0.01~d$^{-1}$ for the near-core rotation rate based on their modelling. 
This seems to match the smallest value of the frequent spacings (i.e. 0.55~d$^{-1}$). The orbital period of the inner binary is 
$\sim$32~d, which implies that the corresponding frequency (at 0.03~d$^{-1}$) would be located in the first bin. {Thus, we 
cannot infer from this plot} whether or not this frequency is affecting the higher {pulsation} frequencies. {
However, the most frequent individual spacing occurs at 0.0012~d$^{-1}$, which is far away from the orbital frequency of the
inner pair.} \\ 
- KIC~8975515 presents an {irregular} distribution of unequal peaks, though of much lower density. The two highest occurrences 
(except for the first bin) 
occur at 2.10 (highest) and 3.76~d$^{-1}$, closely followed by 1.06~d$^{-1}$. Again, the ratio of $\sim$ 2 between two of the three 
most frequent spacings indicates that rotation might be the relevant mechanism to explain the presence of a frequent spacing and its 
harmonic. However, the frequency spacing of 1.66~d$^{-1}$, which probably corresponds to the rotation frequency of the fast-rotating 
hybrid pulsating component in the system \citep{Samadi2020A&A...638A..57S}, is the highest occurrence among the individual spacings.  
\citet{Li2020MNRAS.491.3586L} derived 1.85 $\pm$ 0.01~d$^{-1}$ for the near-core rotation rate. If this rate corresponds to the faster 
rotating component, the core versus surface rotation ratio would equal 1.11 $\pm$ 0.01, as expected from the general trend found by 
\citet{Li2020MNRAS.491.3586L}. 
The frequent spacing of 1.06~d$^{-1}$ might correspond to the more slowly-rotating $\delta$ Scuti component, although 
\citet{Samadi2020A&A...638A..57S} {proposed} a rotation rate of 0.42~d$^{-1}$. In this case, the orbital frequency is 
unresolved. \\
- KIC~9775454 shows a {remarkably} sparse distribution in frequency spacing. The two highest peaks (except for the first bin) 
occur at the spacings of 0.13 and 2.18~d$^{-1}$ (highest). The latter {and most frequent} spacing is accompanied by a few nearby, 
{almost equally frequent} spacings located at 2.07, 2.26 and 2.35~d$^{-1}$. {This suggests} that one of these {frequency 
spacings} is the rotational frequency of the hybrid pulsator (comp~A) (in the case of $\sim$ 2~d$^{-1}$, {it is probably variable}). 
{If we consider the fact that the primary component has \vsini = 65 \kmsi, adopting a default radius of 1.8 R${_\odot}$ corresponding 
to its \teff \citep{Gorda1998ARep...42..793G} and the frequency of 2.18~d$^{-1}$ as the rotational frequency, implies $v_{\mathrm{eq}}$ 
= 199~\kms and a surface inclination angle of $\sim$19\degr~as possible values. The alternative choice (with 0.13 d$^{-1}$) 
would give an impossible value of 12~\kms for $v_{\mathrm{eq}}$.} The orbital frequency is unresolved here as well. \\         
 
\section{Summary and conclusions}\label{sec:conclusions}

In the previous sections, we improved the orbital solutions for four spectroscopic systems from our sample of 49 A/F-type 
{\it Kepler} hybrid stars. 
We improved the orbital solutions for the following spectroscopic systems: KIC~4480321, ~5219533, ~8975515 and KIC~9775454 
by considering the TDs as well as the RVs and performing a simultaneous modelling of both data types. This method allowed 
us to refine the parameters of all the long-period systems, in particular the outer orbits of the triple systems KIC~4480321 
and KIC~5219533. We derived full-fledged SB2 orbital solutions for the long-period binary systems KIC~8975515 and KIC~9775454 
and obtained reliable mass ratios for the first time. Furthermore, the applied methodology enabled us to identify the component 
with the faster $\delta$ Sct-type pulsations, since the TDs are generally linked with the higher frequencies. \\

For KIC~4480321, we concluded that the faster rotating and more massive outer component (comp~C) exhibits the short-period 
$\delta$ Scuti pulsations. Component~C was shown to have a minimum mass of 1.60 $\pm$ 0.05 M$_{\odot}$,  
which can be used as a constraint for asteroseismic modelling. For KIC~5219533, we significantly improved the accuracy of the 
orbital solutions described by \citet{Catanzaro2019MNRAS.487..919C}. We also showed that both orbits are very probably coplanar 
and that the $\delta$ Scuti pulsator is the faster spinning and more massive outer component (comp~C). For KIC~8975515, we 
derived an accurate mass ratio ($q$ = 0.83 $\pm$ 0.05) and established that the faster rotating component (comp~A) exhibits 
$\delta$ Scuti-type pulsations (this concerns the hybrid pulsator). Furthermore, the binary is probably viewed under a low 
inclination angle ($\approx$ 12\degr). The latter is also useful information in an asteroseismic context. For KIC~9775454, 
we determined a precise mass ratio for the first time ($q$ = 0.42 $\pm$ 0.02). We concluded that the more massive component 
of the system (comp~A) is the hybrid pulsating star. \\

From our study of the normalized distributions of the frequency spacings in the high-frequency regime of the Fourier transforms,
we find no firm evidence for the {occurrence} of tidal splitting among the frequent spacings of the {triple} systems with close 
(inner) companions. In the case of the long-period orbits (all systems), it is impossible to resolve the frequency multiplets 
based on the {\it Kepler} data only. On the other hand, {due to the presence of some harmonic frequencies among the most frequent 
spacings}, we {propose the mechanism} of rotational splitting for KIC~4480321 (with a {plausible} surface rotation rate of 
1.63~d$^{-1}$), KIC~5219533 (with a {possible} surface rotation rate of 0.55~d$^{-1}$) and KIC~8975515 (with a {possible} 
surface rotation rate of 1.06~d$^{-1}$ {for the secondary component, whereas the primary rotates at the rate of 1.66~d$^{-1}$)}. 
{In the {particular} case of KIC~9775454, we propose a surface rotation rate of the order of 2~d$^{-1}$, which appears as a 
bunch of closely spaced frequent spacings in the distribution and indicates a possibly variable nature. Detailed analyses of 
the pulsations in the low-frequency regime of the Fourier transforms of these systems, including pattern recognition and period 
spacings of the gravity (and sometimes Rossby) modes for the (rapidly rotating) genuine hybrid pulsators, can provide the 
near-core rotation rates and a confirmation of the proposed surface rotation rates \citep[e.g.,][]{VanReeth2016A&A...593A.120V, 
Li2020MNRAS.491.3586L,Li2020MNRAS.497.4363L}.} \\

\section*{Acknowledgements}

The authors wish to thank the {\it Kepler} team for providing the high-quality data collected with the {\it Kepler} satellite mission. 
They furthermore thank the HERMES Consortium for enabling the production of the high-resolution ground-based spectra, and D.~Bowman et al.\, 
for {authorizing the use of} the additional spectra of KIC~5219533.  
\'AS and ZsB acknowledge financial support of the Lend\"ulet Program \hbox{LP2018-7/2019} of the Hungarian Academy of Sciences. MS acknowledges 
the OV PPP Postdoc\@MUNI grant with nr. \hbox{CZ.02.2.69/0.0/0.0/16\_027/0008360} and the MSMT Inter Transfer program \hbox{LTT20015}. 
ZsB acknowledges the support provided by the National Research, Development and Innovation Fund of Hungary, financed under the PD$_{17}$ 
funding scheme (project PD-123910) and by the J\'anos Bolyai Research Scholarship of the Hungarian Academy of Sciences. HL acknowledges 
support from the grant DFG with nr. \hbox{LE1102/3-1}. ASG acknowledges financial support received from the ALMA-CONICYT grant with nr. 
\hbox{31170029}. {We thank the referee, {Prof. H.~Shibahashi}, for detailed comments and constructive criticism.} We are also grateful 
for support received from the Belgo-Indian Network for Astronomy \& Astrophysics (BINA-1 and BINA-2).
   
\bibliographystyle{aa}
\bibliography{Lampensetal_combined_orbits_v2}


\appendix
\end{document}